\newif\ifisonecolumn
\newcommand{\tr}{{\mbox{tr}}}

\ifisonecolumn
\documentclass[twocolumn,draft,12pt]{IEEEtran}
\else
\documentclass{IEEEtran}
\fi

\ifisonecolumn

\else

\fi

\usepackage{etoolbox}
\usepackage{color}

\usepackage{enumitem}
\usepackage{graphicx}
\usepackage{amsfonts}
\usepackage{amssymb, amsmath}
\usepackage{amsthm}
\usepackage{amsbsy}
\usepackage{amsmath}
\usepackage[english]{babel}
\usepackage{bm}
\usepackage{subfigure} 
\usepackage{cite}
\newtheorem{my_theorem}{Theorem}
\newtheorem{Assumption}{AS}
\newtheorem*{other_theorems}{Theorem}

\newtheorem{my_coro}{Corollary}

\usepackage{epstopdf}
\DeclareMathOperator*{\argmin}{\arg\!\min}
\newcommand{\norm}[1]{\left\lVert#1\right\rVert}
 {
        \theoremstyle{plain}
        
 }
\begin{document}

%+Title
\title{Large Matrix Asymptotic Analysis of ZF and MMSE Crosstalk Cancelers for Wireline Channels}
\author{Itsik Bergel,~\IEEEmembership{Senior Member,~IEEE}, S.~M.~Zafaruddin,~\IEEEmembership{Member,~IEEE} 
        and Amir~Leshem,~\IEEEmembership{Senior Member,~IEEE}%
        \thanks{The authors are with the Faculty of Engineering, Bar-Ilan University,
                Ramat Gan 52900, Israel (e-mail: itsik.bergel@biu.ac.il, smzafar@biu.ac.il, leshema@biu.ac.il). S. M. Zafaruddin was partially funded by the Israeli Planning
                and Budget Committee (PBC) post-doctoral fellowship. A portion of this paper was presented at the IEEE ICSEE 2016 \cite{zafaruddin_eilat2016}.}% <-this % stops a space\maketitle
}

\maketitle
%-Title

\begin{abstract}
 We present asymptotic expressions for user throughput in a multi-user wireline system with a linear decoder, in increasingly large system sizes. This analysis can be seen as a generalization of results obtained for wireless communication. The  features of the diagonal elements of the wireline channel matrices make wireless asymptotic analyses inapplicable for wireline systems. Further, direct application of results from random matrix theory (RMT) yields a trivial lower bound.   This paper presents a novel approach to asymptotic  analysis,  where an alternative sequence of systems is constructed that includes the system of interest in order to approximate  the spectral efficiency of the linear zero-forcing (ZF) and minimum mean squared error (MMSE) crosstalk cancelers.  Using  works in the field of large dimensional random matrices, we show that the user rate in this sequence converges to a non-zero rate. The approximation of the user rate for both the ZF and MMSE cancelers are very simple to evaluate and does not need to take specific channel realizations into account. The analysis reveals the intricate behavior of the throughput as a function of the transmission power and the channel crosstalk. This unique behavior has not been observed  for linear decoders in other systems. The  approximation presented here is much more useful for the next generation   G.fast wireline system than earlier digital subscriber line (DSL) systems as previously computed performance bounds, which are strictly larger than zero only  at low frequencies. We also provide a numerical performance analysis  over measured and simulated DSL channels which show that the  approximation is accurate even for relatively low dimensional systems and is useful for many  scenarios in practical DSL systems.
\end{abstract}

\section{Introduction}
Wireline digital subscriber line (DSL) systems use the existing infrastructure of telephone networks 
to provide broadband services to customers \cite{starr03}. The new G.fast standard targets fiber-like speed (e.g., upto $1$ Gbps) over short copper loops \cite{g9701}, \cite{timmers2013mag}, \cite{zafar2017spm}. Until the final migration of all copper lines to fiber in access networks,  DSL technology is  likely to continue to be used widely,  and play a key role in  the convergence of next generation wired and wireless technologies. Another wireline technology  is the  10GBASE-T Ethernet to provide a  data rate of $10$ \mbox{Gb/s} over structured copper cabling systems. However,  the performance of  wireline  systems is limited by interference caused by the electromagnetic coupling of transmissions from other wire pairs. Specifically, the main issue is far-end crosstalk (FEXT), which is generated by the transmitters at the opposite side of the binder \cite{starr03}.

With the advent of vectoring technology  \cite{ginis2002vectored}, \cite{g9935},  various techniques  have been developed to cancel crosstalk in  multichannel wireline systems \cite{zafar2017spm}. 
These include  linear   zero-forcing (ZF) \cite{cendrillon2006near}, \cite{cendrillon2007}  and linear minimum mean squared error (MMSE) equalizers \cite{wahibi08hindawi}, \cite{pandey2010mmse}, and  non-linear based techniques such as the ZF generalized decision feedback equalizer (ZF-GDFE) \cite{chen2007optimized} and  MMSE-GDFE \cite{Tsiaflakis2007OSB}.  Cendrillon \emph{et al.} \cite{cendrillon2006near} \cite{cendrillon2007} showed that 
the ZF processing is close to optimal in typical DSL channels due to the diagonal dominance of the channel matrix. Thus, over the years, the ZF method has become very popular for upstream  decoding
 (e.g., \cite{chen2007optimized}\nocite{zafaruddin2012performance}-\cite{bergel2013performance}), and  downstream precoding (e.g., \cite{leshem2007low}\nocite{bergel2010convergence}\nocite{binyamini2013adaptive}\nocite{strobel2015icc}\nocite{strobel2015gc}-\cite{Barthelme2016gc}) in DSL systems. However, with increasing frequency, diagonal dominance declines  and the MMSE canceler outperforms the ZF canceler particularly at higher frequency tones.  While more advanced non-linear receivers have been proposed \cite{ginis2002vectored}, \cite{chen2007optimized}, these simple receivers tend to be preferred, particularly in the computationally intensive crosstalk cancellation of DSL systems. The G.fast standard recommends the use of linear structures for crosstalk cancellation for $106$ \mbox{MHz} \cite{g9701}.

The performance of the ZF canceler had been shown to be dependent on the diagonal dominance characteristics of the DSL channel \cite{cendrillon2006near, cendrillon2007, zafaruddin2012performance, bergel2013performance}.  Newer performance bounds \cite{zafaruddin2012performance, bergel2013performance} were shown to be even tighter,  and guaranteed the ZF near optimally for even higher frequencies. The bounds in \cite{bergel2013performance}  are much simpler to evaluate, and better show the near optimality of ZF processing when the
channel matrix is diagonally dominant.  However, with the increased bandwidth of G.fast, the FEXT is higher and all of these bounds in \cite{cendrillon2006near, cendrillon2007, zafaruddin2012performance, bergel2013performance} become irrelevant (they degenerate to a lower bound of $0$).    Surprisingly, in many cases the ZF canceler still performs well, and in other cases the MMSE canceler is close to optimal.  However, no analysis has been conducted to confirm these outcomes.

In this work, we examine the asymptotic behavior of ZF  and MMSE decoders for wireline channels as the number of jointly decoded users grows, using tools from the field of large dimensional random matrix theory (RMT). Most works on RMT have dealt with the case of zero-mean random matrices, and have investigated the  asymptote of the spectrum of $N \times N$ random matrices $\mathbf Z_n \mathbf Z_n^H$ where $\mathbf Z_n$ has zero mean entries. For example, Marchenko and Pastur \cite{marchenko1967distribution} and Silverstein and Bai  \cite{silverstein1995empirical} analyzed these matrices with i.i.d. entries, whereas Girko  \cite{girko2001theory} and Khorunzhy et al.  \cite{khorunzhy1996asymptotic} discussed these matrices with non-i.i.d. entries. The case of non-zero mean random matrices has been less widely explored. Nevertheless, Dozier and Silverstein \cite{dozier2007empirical} and Hachem et al. \cite{hachem2007deterministic} presented a deterministic equivalent of the empirical Stieltjes transform of $\mathbf Z_n \mathbf Z_n^H$, where $\mathbf Z_n = \mathbf Y_n+ \mathbf A_n$ with $\mathbf Y_n$ as a zero mean random matrix and $\mathbf A_n$ as a sequence of deterministic matrices.

The use  of large RMT enables a deterministic analysis of the performance of systems that are by nature random and quite complex. This approach has been applied to analyze the performance of linear decoders for wireless networks \cite{Tse1999,Tse2000, Liang2007_mmse, Kumar2009, Hoydis2010}. In particular,   Liang et al. \cite{Liang2007_mmse} used the results of \cite{girko2001theory} for asymptotic analysis of the MMSE performance in MIMO wireless systems whose channel coefficients are all identically distributed. However, the wireline system differs from the wireless system, and thus analysis methods for MIMO MMSE/ZF can not  be readily applied  in the wireline context.  This is primarily because the diagonal elements in the wireline channel matrix are  different from the non-diagonal elements, which is not the case in  wireless systems. Thus,  a novel approach is required to get useful results in wireline systems. Moreover, the RMT of  \cite{hachem2007deterministic}, although applicable to wireline systems cannot be directly applied without adaptation to the parameters of wireline systems.
 
In this paper, we take a novel approach to the asymptotic analysis of large systems where we construct an alternative sequence of systems that includes the system of interest. We then use the results in \cite{hachem2007deterministic}  to derive an approximation of the spectral efficiency of linear ZF and MMSE decoders for wireline systems. We also provide numerical results over measured and simulated DSL channel matrices to demonstrate the accuracy of the analysis for various system parameters.  

The presented approximation is very simple to evaluate and does not require the knowledge of the  specific channel.  The performance of the ZF decoder is shown to decrease linearly  with  FEXT power up to the point where the FEXT power is equal to the direct channel power. If the FEXT power is larger than the direct channel power, the asymptotic performance of the ZF decoder approaches zero. 

The MMSE decoder exhibits a more intricate behavior. For example, the asymptotic behavior of the output SNR for high input power can have three different behaviors: it can be proportional to the power $P$, proportional to the square root of the power $\sqrt{P}$, or proportional to  $P^{2/3}$ depending on the average FEXT power. The behavior of the MMSE SNR as a function of the FEXT power is also non-trivial. It can be either monotonically increasing or can have a local minimum at a FEXT power that is less than twice the direct channel power, depending on the input power. This intricate behavior is very different than the performance of MMSE decoders for any other scenario.

 Note that the proposed approximation is much more optimistic than existing deterministic analyses \cite{cendrillon2006near}, \cite{chen2007optimized}, \cite{bergel2013performance} which placed limits on the sum of the absolute values of the FEXT terms.  

We also show that the proposed asymptotic analysis  degenerates into the wireless solution \cite{Liang2007_mmse}, with the proper choice of  system parameters when the variables are indeed i.i.d. Thus, the  analysis can be seen as a generalization of the \cite{Liang2007_mmse} to a more general setup that includes the wireline scenario.

The rest of this paper is organized as follows. Section II defines the wireline system and channel models for DSL system. The novel asymptotic analysis approach is described in Section III.  The performance of ZF and MMSE cancelers is described Section IV.  Section V provides the performance evaluation using numerical analysis on measured and simulated channel data of DSL systems. Section VI concludes the paper.

\section{System Model}
\subsection{System Setup}
The multi-user wireline channel is modeled as a multiple-input multiple-output (MIMO) system (known as vectored system in DSL technology) with $M$ users that are connected to the distribution point (DP) through a cable of $M$ twisted pairs \cite{zafar2017spm}. We consider upstream transmission and assume perfect synchronization among users. Discrete multi-tone (DMT) modulation  is used that facilitates  independent processing at each tone.
The signal vector $\mathbf y\in \mathbb{C}^{M}$ received by the DP at any given symbol time and any given frequency tone can be written as: 
\begin{equation}
\mathbf y = \sqrt{p} \mathbf H_\mathrm c \mathbf x+ \mathbf v
\end{equation}
where $\mathbf H_\mathrm c \in \mathbb{C}^{M\times M}$ is the channel matrix in which the $i,j$ element represents the channel coefficient from user $j$ to the ports of the $i$-th pair in the DP, $\mathbf x \in \mathbb{C}^{M}$  is a vector that contains the transmitted symbols of all users, $\mathbf {v} \in \mathbb {C}^{M}$ is a complex Gaussian noise with zero mean and variance $\sigma^2_{v}$, and $p$ is the transmitted power at the given frequency tone.
Without loss of generality, we assume that all transmitted symbols (i.e., the elements of the vector $\bf x$) are independent and identically distributed (i.i.d.) Gaussian random variables with zero mean and unit variance, and that the transmission powers of all users are equal. 
%Hence, it is of interest to characterize the statistical distribution of the achievable rates over the ensemble of all binders of the same type. 
\subsection{Wireline Channel Matrix $\mathbf{H}_\mathrm c$}
The diagonal elements of the channel matrix $\mathbf{H}_\mathrm c$ represents the attenuation of the direct signals while the off-diagonal elements reflect the crosstalk.
The performance of a wireline system depends on this matrix of channel gains, which can be measured for a specific binder under specific environmental conditions. It is known that the diagonal part of a DSL channel is a function of frequency, loop length and physical parameters of the twisted pairs \cite{starr03}. However, the variation of the gain of these direct channels between different wires with the same parameters is relatively small such that this direct channel gain is often considered to be  deterministic. However, the off-diagonal elements depend on various other factors such as  capacitance and the inductive imbalance between the pairs, non-uniform twisting,  geometric imperfections of twisted  pairs.  These lead to relatively large variations in the crosstalk couplings. Hence,  non-diagonal elements of the DSL channel matrix are often considered to be random \cite{lin80a, U-BROAD, maes09, Brink2017}.  

To analyze the performance of wireline systems, we need to  know its channel matrix.  However, such measurements  are rarely available in advance and do not cover a wide range of scenarios. As a substitute, one can turn to statistical analysis. Statistical models are available and have been presented by Karipidis et al. \cite{karipidis2006crosstalk} and Sorbara  et al. \cite{sorbara2007construction} \cite{Brink2017} for DSL systems. These models have been adopted by various studies of DSL systems (e.g., \cite{zafaruddin2012performance},  \cite{baldi2010simple}, \cite{gomaa2011new}).

A widely acceptable   statistical model of  the DSL  FEXT coupling \cite{sorbara2007construction}   ${[H_c]}_{ij}$ , $i\neq j$ is given by
\begin{align}
{[H_c]}_{ij}=K_{\rm fext}{[H_c]}_{jj}f\sqrt{l_{ij}}C_{ij}, \forall i\neq j 
\label{eq:crosstalk_model}
\end{align}
where $f$ is frequency of operation,  $l_{ij}$ is coupling length, and $K_{\rm fext}$ is a constant that depends on the type of cable (e.g. for $24$ AWG cables  $K_{\rm fext}= 1.59 \times 10^{-10}$). The term ${[H_c]}_{ij}$ denotes the direct path of the disturber. The dispersion (excluding the phase) is modeled by a log-normal random variable,
$C_{ij}=10^{-0.05\chi(f)}\exp({j\phi})$ where $\chi(f)$  is a Gaussian random variable in dB with mean $\mu_{\text{dB}}=2.33\,\sigma_{\text{dB}}$ and variance $\sigma_{\text{dB}}$, and $\phi$ is uniform phase in the interval [$0, 2\pi$]. The model of the direct path is given as:
\begin{align}
{[H_c]}_{jj}=\exp(-lr)
\label{eq:direct_model}
\end{align}
where $l$ is line length of the $j$-th user and $r$ is the attenuation constant of the cable. Extensive measurement campaigns are used to derive these parametric cable models for the diagonal and non-diagonal elements of the channel matrix. As seen in (\ref{eq:crosstalk_model}) and (\ref{eq:direct_model}), the non-diagonal and diagonally are distributed differently. 

 Since the randomness of the DSL channel is exhibited in the non-diagonal elements and not in the diagonal part, we define a normalized FEXT matrix $\mathbf Q$ whose elements are 
\begin{align}
\begin{split}
q_{ij}& ={[H_c]}_{ij}/{[H_c]}_{jj},~~ \forall i\neq j\\
       &=0,~~~~~~~~~~~~~~ \forall i= j.
\end{split}
\end{align}

Thus, the channel matrix can be decomposed as:
\begin{equation}\label{d: Q}
\mathbf H=\mathbf I+\mathbf Q, ~~~
\mathbf H_\mathrm c= \mathbf H \mathbf D 
\end{equation}
where $\mathbf D \in \mathbb{C}^{M\times M}$ is a diagonal matrix with the diagonal elements of $\mathbf {H}_c$ (thus,  $\mathbf H\in \mathbb{C}^{M\times M}$ is the normalized channel matrix, in which all diagonal elements are equal to $1$).

Our asymptotic performance analysis does not rely on the specifics of a particular model,  and requires only the channel structure of (\ref{d: Q}) and the following assumption:
\begin{Assumption}\label{assump:one}
The matrix $\mathbf Q$ is statistically independent of the matrix $\mathbf D$.
\end{Assumption}

\begin{Assumption}\label{assump:two}
        All (off-diagonal) elements of $\mathbf Q$ are identically distributed and statistically independent (i.i.d.).
\end{Assumption}
An examination of  the parametric models in \eqref{eq:crosstalk_model} and \eqref{eq:direct_model} shows the elements of $\mathbf Q$ are i.i.d in the case of an equal length binder.  Note that these  assumptions are less restrictive, and can also accommodate other channel models. In \cite{zafaruddin_eilat2016}, we  studied the statistical characterization  of a DSL channel and verified these assumptions using measured data.

\section{Asymptotic Analysis}\label{sec:performance_analysis}

\subsection{Bounds on the Average Spectral Efficiency }

In this sub-section, we evaluate the average  spectral efficiency for the ZF and MMSE cancelers and derive a convenient lower bound. This analysis will be used to derive the asymptotic analysis in the next sub-section. 

In order to extract the transmitted signals, the receiver multiplies the received signal by a linear equalizer matrix $\mathbf F$ so that the estimate of the vector $\mathbf x$ is:
\begin{equation} 
\hat{\mathbf x} =\mathbf F \mathbf y.
\label{eq:est_x}    
\end{equation}

The resulting signal to interference plus noise ratio
(SINR) for the $i$-th is:
\begin{equation}
\rho_i=\frac{p|[\mathbf {FH}_c]_{i,i}|^2}{p\sum_{j\ne i}|[\mathbf{FH}_c]_{i,j}|^2+\sigma_v^2 [\mathbf{F F}^H]_{i,i}}
\end{equation}

and the resulting spectral efficiency is 
\begin{equation}
R_i=\log_2(1+\rho_i).
\end{equation}

For reference we also define the single wire (SW) performance,  if only one user transmits and the DP only uses  the active wire for  detection. The single wire rate is:
\begin{equation} \label{eq:spectral_awgn}
R_i^{\rm }= {\log_2(1+\eta_i)},
\end{equation}
where $\eta_i=\frac{p|d_{i,i}|^2}{\sigma_ v^2}$ is the SW-SNR of the $i$-th user, and  $d_{i,i}$ is the $i$-th element on the diagonal of  matrix $\mathbf{D}$.

\subsubsection{ZF Linear Canceler}
For the ZF, the equalizer matrix is given as $ \mathbf F=\mathbf H^{-1}=\mathbf D \mathbf H_\mathrm c^{-1} $ which converts \eqref{eq:est_x}  to $\hat{\mathbf x} =\sqrt{p}\mathbf D \mathbf x+\mathbf  H^{-1} \mathbf v$.
The SINR of user $i$ given the channel matrix, $\mathbf{H}_\mathrm c$  is given by:
\begin{eqnarray} \label{eq:sinr_zf}
\rho_i=\frac{p \mathbb E[|d_{i,i}x_{i}|^2|d_{i,i}]}{ \mathbb E[|(\mathbf{H}^{-1} \mathbf v)_i|^2|\mathbf{H}]}=\frac{\eta_i}{((\mathbf H^H \mathbf H)^{-1})_{i,i}} 
\end{eqnarray}
The average spectral efficiency of user $i$ is given by:
\begin{equation} \label{eq:rr}
R_i=\mathbb E\bigg[{\log\Big(1+\frac{\eta_i}{((\mathbf H^H \mathbf H)^{-1})_{i,i}}\Big)}\bigg]
\end{equation} 
where the expectation is taken with respect to the distribution of the channel matrix.
By comparing \eqref{eq:spectral_awgn} to (\ref{eq:rr}), we define the SNR loss with the ZF canceler as $\gamma_i=((\mathbf H^H \mathbf H)^{-1})_{i,i}$ such that the spectral efficiency becomes:
\begin{equation} \label{eq:rr2}
R_i=\mathbb E\bigg[{\log\Big(1+\frac{\eta_i}{\gamma_i}\Big)}\bigg].
\end{equation} 

 Using   Jensen's inequality, the spectral efficiency is lower bounded by:
\begin{equation} \label{eq:r_lb1}
R_i \geq \tilde{R_i}=\mathbb E\bigg[\log\bigg(1+\frac{\eta_i}{\mathbb E[\gamma_i]}\bigg)\bigg ].
\end{equation} 

Using   assumptions AS\ref{assump:one} and AS\ref{assump:two}, the distribution of $\gamma_i$ is identical for all users. Hence
\begin{equation}
\mathbb E[\gamma_i]=\frac{1}{M}\sum_{i=1}^M \mathbb E[\gamma_i]=\mathbb E[\bar \gamma]
\label{eq: gamma avr3}
\end{equation}
where 
\begin{align} 
\label{eq:zf:trace}
\begin{split}
\bar {\gamma}& =\frac{1}{M}\sum_{j=1}^M\gamma_j =\frac{1}{M}\sum_{i=1}^M\Big[(( \mathbf H^H \mathbf H)^{-1})_{i,i}\Big]\\&=\frac{1}{M}\tr \left[\big[( \mathbf H^H \mathbf H)^{-1}\big]\right]
\end{split}
\end{align}
where $\tr[\mathbf \cdot]$ denotes the trace of a matrix. 

The substitution of (\ref{eq: gamma avr3}) into (\ref{eq:r_lb1})  significantly simplifies the bound, and also allows us to apply asymptotic results.
\subsubsection{MMSE Linear Canceler}

For the MMSE,  the canceler matrix can be obtained using $\argmin_{\mathbf F} \mathbb{E}[\norm{\mathbf{\sqrt{p}Dx}-\mathbf{Fy}}^2]$ to get\footnote{Note that the multiplication on the left by $p|\mathbf{D}|^2$ is not required in a detection setup since it has no effect on the detection of the SINR.}:
\begin{align}\label{eq:F_MMSE}
\mathbf F=  {p}|\mathbf{D}|^{2}\mathbf{H}^H(p\mathbf{H}|\mathbf{D}|^2\mathbf{H}^H+\sigma_v^2\mathbf{I})^{-1}
\end{align} 
where $|\mathbf{D}|^2$ is the square of absolute values of the elements of  matrix $\mathbf{D}$.
The error covariance matrix for the estimation of $\mathbf{x}$ is:
\begin{eqnarray} \label{eq:error_cov_mmse}
\begin{split}
\mathbf C_{e}&=\mathbb{E}\{(\mathbf{\sqrt{p}Dx}-\mathbf{F}\mathbf{y})(\mathbf{\sqrt{p}Dx}-\mathbf{F}\mathbf{y})^H\}\\&= p|\mathbf{D}|^2-{p}|\mathbf{D}|^{2}\mathbf{H}^H(p\mathbf{H}|\mathbf{D}|^2\mathbf{H}^H+\sigma_v^2\mathbf{I})^{-1}p\mathbf{H}|\mathbf{D}|^2\\
&= \sigma_v^2\left[\mathbf{H}^H\mathbf{H}+\frac{\sigma_v^2}{p}|\mathbf{D}|^{-2}\right]^{-1}
\end{split}
\end{eqnarray}
where we used the matrix inversion lemma:   $(\mathbf A+\mathbf B \mathbf C)^{-1} = \mathbf A^{-1}-\mathbf A^{-1} \mathbf B( \mathbf{I}+ \mathbf{C}\mathbf A^{-1} \mathbf B)^{-1}\mathbf C\mathbf A^{-1}$. Manipulating \eqref{eq:F_MMSE} and \eqref{eq:error_cov_mmse} leads to the well known result:
\begin{eqnarray} \label{eq:sinr_mmse}
\rho_i= \frac{p\left[|\mathbf{D}|^2\right]_{ii}}{[\mathbf {C}_e]_{ii}}-1 
\end{eqnarray}
Thus average spectral efficiency is:
\begin{equation} \label{eq:spectral_efficiency}
R_i=\mathbb E\bigg[\log \big(\frac{\frac{p}{\sigma_v^2}\left[|\mathbf{D}|^2\right]_{ii}}{\gamma_i}\big)\bigg],
\end{equation} 
where $ {\gamma}_i = (\mathbf{H}^H\mathbf{H}+\frac{\sigma_v^2}{p}|\mathbf{D}|^{-2})^{-1}_{ii}$. As the diagonal matrix $\mathbf{D}$ is deterministic, we can apply Jensen's inequality again as in the ZF case to bound \eqref{eq:spectral_efficiency} by:
\begin{equation} \label{eq:spectral_efficiency2}
R_i\geq\log \big(\frac{\frac{p}{\sigma_v^2}\left[|\mathbf{D}|^2\right]_{ii}}{\bar{\gamma}}\big).
\end{equation} 
where $\bar {\gamma} = \frac{1}{M}\sum_{j=1}^M\gamma_j$. Note that in the MMSE canceler case, the average SNR loss, i.e., the ratio between SW-SNR and   the SNR at output of the MMSE canceler is given by:$ \frac{\eta_i \bar {\gamma}}{\eta_i-\bar {\gamma}}$. 
Using  matrix notation, the $\bar {\gamma}$-parameter of the MMSE canceler can be simplified:
\begin{eqnarray} \label{eq:gamma_mmse}
\bar {\gamma} =\frac{1}{M} \text{tr} [(\mathbf{H}^H\mathbf{H}+\frac{\sigma_v^2}{p}|\mathbf{D}|^{-2})^{-1}].
\end{eqnarray}

In the case that all diagonal elements of $\mathbf D$ are equal i.e., $d_{ii} =d~~ \forall i$ such that $\mathbf D =d \mathbf{I}$ and using  $\eta = \frac{|d|^2p}{ \sigma_v^2}$  in (\ref{eq:gamma_mmse}), we get

\begin{eqnarray} \label{eq:gamma_general1}
\bar {\gamma} =\frac{1}{M} \text{tr} [(\mathbf{H}^H\mathbf{H}+\frac{1}{\eta}\mathbf{I}_M)^{-1}].
\end{eqnarray}
We observe that both  performance metrics can be represented as:
\begin{eqnarray} \label{eq:gamma_general}
\bar {\gamma} =\frac{1}{M} \text{tr} [(\mathbf{H}^H\mathbf{H}+\frac{1}{\xi}\mathbf{I}_M)^{-1}].
\end{eqnarray}
where for MMSE $\xi=\eta$ and for ZF $\xi\to \infty$. Representation of the performance metrics in the form of (\ref{eq:gamma_general}) allows us to apply the large matrix result of Hachem et al. \cite{hachem2007deterministic}.

In this work, we perform an asymptotic analysis of the $\bar{\gamma}$-parameter, and derive simple expressions that do not require the knowledge of the specific channel realizations.  We show that in the asymptotic regime the performance of the ZF and MMSE cancelers  converges to a constant and their average spectral efficiencies in \eqref{eq:rr} and \eqref{eq:spectral_efficiency}  become tight.

\subsection{Novel  Asymptotic Analysis Approach } \label{subsec:Asymptotic_Analysis}
In this section, we derive an approximation to the  $\bar {\gamma}$-parameter for both the ZF and MMSE cancelers using the method of large matrix analysis. Before we start the asymptotic analysis, we need to note that the traditional analysis approach as the system size grows to infinity ($M\rightarrow\infty$) does not lead to useful limits. In the ZF case, such an asymptotic analysis converges to a zero rate, which cannot give a reasonable approximation for the performance. In the MMSE case, the rate converges to a non-zero limit, but in a way that cannot account for the importance of the direct channel elements (the diagonal elements in the channel matrix).  Instead, we present a novel approach in which we construct a new sequence of systems such that the system of interest (described in the previous section) is an element in the sequence, and the rate of each user in this sequence of systems converges to a non-zero limit.

To derive this new sequence of systems, we define:
\begin{equation} \label{eq:Sigma}
\mathbf \Sigma_n = \mathbf I_n + \sqrt{\frac{M-1}{n-1}}\mathbf Q_n
\end{equation}
 a sequence of matrices with increasing sizes, where $n \ge2$ and $\mathbf Q_n$ is an $n \times n$ random matrix with zero diagonal and i.i.d. elements outside the diagonal. We set the distribution of each non-diagonal element of $\mathbf Q_n$ ($q_{n,i,j}$ for $i\ne j$) to be identical to the distribution of a non-diagonal element in $\mathbf Q$, and require that $\mathbb{E}\left[| q_{1,2}|^{4+\epsilon}\right]$  be bounded for some $\epsilon>0$.  Note that $\mathbf \Sigma_n$ has the same distribution as $\bf H$.
Thus, the definition of $\mathbf \Sigma_n$ establishes the sequence of arbitrary matrices which intersect with our system model when $n=M$. More specifically, this sequence is constructed such  that the total FEXT power per user is constant for all system sizes. 

\begin{figure}[t]
        \centering
        {\includegraphics[scale=0.45]{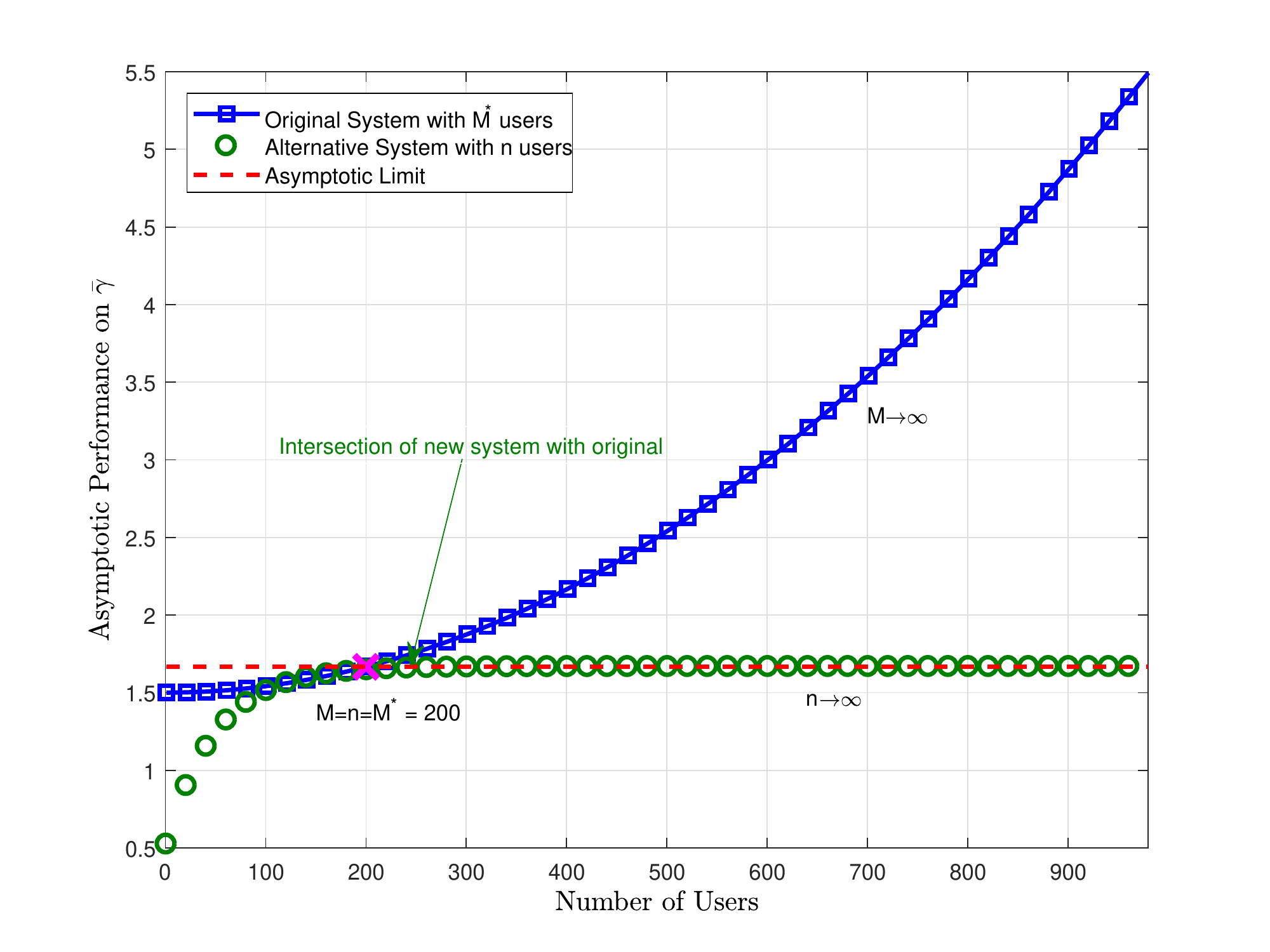}}
        \caption{An example of the construction of the sequence of systems for asymptotic analysis.}
        \label{fig:asym_seq}
\end{figure}

This novel asymptotic analysis approach is illustrated in Fig.~\ref{fig:asym_seq}, which depicts the SNR loss of a ZF system. The blue squares show that the SNR loss diverges as the system size, $M$,  grows to infinity (and hence the rate converges to zero). On the other hand, the alternative sequence (depicted by circles) converges to a finite bound (depicted by the dashed line). The original and the alternative sequences intersect at the size of interest $n=M=200$. Clearly the dashed line  gives a good approximation for the SNR loss of both sequences at this point. 

Using \eqref{eq:gamma_general}, the associated  $\bar{\gamma}$-parameter for the alternative sequence system of size $n$ is represented as 
\begin{align} \label{eq:both_zf_mmse}
\tilde {\gamma}_n&=
\begin{cases}
\frac{1}{n}\tr\left[(\mathbf \Sigma^H_n \mathbf \Sigma_n+\frac{1}{\xi}\mathbf{I})^{-1} \right].
\end{cases}
\end{align}
These two new quantities, $\mathbf \Sigma_n$ and $\tilde {\gamma}_n$, will be used in the following to conduct  the asymptotic analysis as  $n \to \infty$. The asymptotic analysis results in a deterministic equivalent for both ZF and MMSE cancelers, which describes the behavior of the lower bound as the system size becomes large enough. 

It is important to note that practically speaking, our approach has exactly the same meaning as any other asymptotic analysis. That is, if the system of interest is far enough along in the sequence, its performance is well approximated by the asymptotic results. Also note that in most types of asymptotic analysis, the important question: ``How close
is the system of interest to the asymptotic result?" is primarily dealt with through simulations (for example in large matrix analysis \cite{chuah2002capacity}--\nocite{cardoso2008cooperative}\cite{marzetta2010noncooperative}, and in asymptotic SNR analyses using `degrees of freedom' \cite{zheng2003diversity}--\nocite{jafar2008degrees}\cite{jung2012opportunistic}). This question  will be addressed below by an analysis using some special cases and extensively through numerical and simulation examples in Section \ref{Sec:Numerical_s}. 

For the asymptotic analysis, we use the theorem presented in Hachem et al. for non-centralized large random matrices \cite{hachem2007deterministic}. 
To use the result, we  adapt our parameters to the parameters in \cite{hachem2007deterministic} by setting $\sigma_{i,j}^2(n)=\frac{n}{n-1}(M-1) \mathbb E[|q_{n,i,j}|^2]$, and $x_{i,j}=\frac{q_{n,i,j}}{\sqrt{\mathbb{E}[|q_{n,i,j}|^2]|_{i\ne j}}}$. Hence, the theorem presented in Hachem et al. \cite{hachem2007deterministic}  can be represented in the following Theorem for our system setup.

\begin{other_theorems}[Hachem et al. \cite{hachem2007deterministic}] 
        Consider an $n \times n$ random matrix $\mathbf Y$ in which the $i,j$ entry is given by $y_{i,j}=\frac{{\sigma_{i,j}(n)}}{\sqrt{n}}x_{i,j}$, and $x_{i,j}$ are i.i.d. random variables with zero mean and unit variance, which  satisfy the following assumptions: 
        
        (1) There exists an $\varepsilon > 0$ such that: $\mathbb E\left[|x_{i,j}|^{4+\epsilon}\right]<\infty $ 
        
        (2) There exists a finite number $\sigma_{max}$ such that: $ \sup_{n \ge 1} \max_{i,j}|{\sigma_{i,j}(n)}| \le \sigma_{max}$
        
        Let $\mathbf{\Sigma = I + Y}$.  There exists a deterministic $n \times n$ matrix-valued function $\mathbf T(z)$ analytic in $\mathbb{C}-\mathbb{R}^+$ such that, almost surely:
        \begin{equation} \label{eq:main_th} 
        \lim_{n\to \infty}\left( \frac{1}{n}tr[( \mathbf \Sigma \mathbf \Sigma^H-z\mathbf I_n)^{-1}] - \frac{1}{n}tr [\mathbf T(z)]\right)=0
        \end{equation}
        and $\mathbf T(z)$ is the unique positive solution of:
        \begin{align}
        \begin{split} 
        \mathbf T(z)=\left(\mathbf \Psi^{-1}(z)-z  \tilde{\mathbf \Psi}(z) \right)^{-1} \\
        \tilde{\mathbf T}(z)=\left(\tilde{\mathbf \Psi}^{-1}(z)-z  \mathbf \Psi(z) \right)^{-1}    
        \label{eq:T(z)}
        \end{split} 
        \end{align}
        \begin{equation}\label{eq:Psi(z)}
        \begin{split}
        \mathbf \Psi(z)=diag(\psi_1(z),\ldots,\psi_n(z)) \\
        \tilde{\mathbf \Psi}(z)=diag(\tilde{\psi}_1(z),\ldots,\tilde{\psi}_n(z))
        \end{split}
        \end{equation}
        \begin{equation} \label{eq:psi(z)}
        \begin{split}
        \psi_i(z)=-z^{-1} (1+ tr(\tilde{\mathbf \Omega}_i\tilde{\mathbf T}(z))/n )^{-1} \\
        \tilde{\psi}_j(z)=-z^{-1}(1+tr({\mathbf \Omega_j}{\mathbf T}(z))/n )^{-1}
        \end{split} 
        \end{equation}
        \begin{equation} \label{eq:D_i}
        \begin{split}
        &\tilde{\mathbf \Omega}_i=diag(\sigma_{i1}^2(n),\ldots,\sigma_{in}^2(n))\\
        &{\mathbf \Omega}_j=diag(\sigma_{1j}^2(n),\ldots,\sigma_{nj}^2(n)).
        \end{split}
        \end{equation}
        \label{theorem1}
\end{other_theorems} 
The main rationale for using  Theorem 1 is that the performance of linear cancelers can be represented as a deterministic function that does not depend on the actual channel realization. In addition, the solution in \eqref{eq:main_th}-\eqref{eq:D_i} only involves diagonal matrices in contrast to the  matrix inversion required in the original problem. While the result of this theorem is still quite complicated, we will show that  we can simplify the equations to a closed form performance expression.

As mentioned above, in this work we cannot directly apply the Hachem theorem to the problem at hand. Instead we use our alternative   choice of  $\mathbf{\Sigma}$ which enables accurate approximation of the  parameters. In the next subsection, we derive our main theorem by applying the Hachem theorem to the new sequence of systems, defined in \eqref{eq:Sigma}.  We then further use the system properties to derive a simple expressions for our performance, characterized by $\gamma^o$.

\subsection{Asymptotic Analysis Theorems}
\begin{my_theorem}\label{theorem:main}
        Let
\begin{align}\label{eq:sigma2}
\begin{split}
\sigma^2=(M-1)\mathbb E[|q_{i,j}|^2]\big|_{i\ne j}
\end{split}
\end{align}
be the total average FEXT power, then the deterministic equivalent  $\gamma_{\rm }^{\circ}$ is the unique real and positive  root of the cubic equation in $t$
        \begin{align}\label{eq:cubic_main}
        \sigma^4t^3     +2\sigma^2 t^2+ (1+\xi-\xi \sigma^2)t-\xi =0.
        \end{align}
        
                As the system size grows to infinity,  $\tilde{\gamma}_n$-parameter defined in \eqref{eq:both_zf_mmse} will converge almost surely to the deterministic equivalent:
        \begin{equation} \label{eq:mmse_th}
        \lim_{n\to \infty} (\gamma_{\rm}^{\circ}-\tilde\gamma_n)=0 .
        \end{equation}
        
\end{my_theorem}

\begin{IEEEproof}
Comparing \eqref{eq:main_th}  with \eqref{eq:both_zf_mmse}   and noting that the matrix $\mathbf {\Sigma \Sigma}^H$ is full rank, asymptotic performance  on the $\tilde{\gamma}_n$ for the asymptotic analysis can be obtained  by evaluating
\eqref{eq:main_th}:
\begin{equation}\label{eq:mmse_asymp}
\lim_{n\to \infty}\tilde \gamma_n =\gamma_{\rm}^{\circ} \triangleq  \lim_{n\to\infty,} \frac{1}{n}\tr[\mathbf T(-\frac{1}{\xi})].
\end{equation}

To evaluate \eqref{eq:mmse_asymp}, we simplify equations (\ref{eq:main_th})-(\ref{eq:D_i}). Due to the homogeneity of the variance matrix, $\Sigma_n$ (except for its diagonal), we have $\mathbf \Omega_i=\tilde{\mathbf \Omega}_i$ and the $j$-th element on the diagonal of $\mathbf \Omega_i$ equals $\frac{n}{n-1}\sigma^2$ if $j\ne i$ and $0$ if $j=i$. Since $\mathbf T(-\frac{1}{\xi})$ is a diagonal matrix, we define $t_i=(\mathbf T(-\frac{1}{\xi}))_{i,i}$ and $t=\tr(\mathbf T(-\frac{1}{\xi}))/n$, which yields: 
        \begin{IEEEeqnarray}{rCl}
                \tr({\mathbf \Omega_j}{\mathbf T}(-\frac{1}{\xi}))= \frac{n}{n-1}\sigma^2 (n t-t_i).
        \end{IEEEeqnarray}
        
        Using the same definitions for $\tilde t_i$ and $\tilde t$, equation (\ref{eq:T(z)}) becomes:
        \begin{align}\label{eq: t sec}
                \begin{split} 
                        &t_i=\frac{1}{ \frac{1}{\xi}\left (1+ \frac{\sigma^2}{n-1} (nt-t_i\right) +\frac{1}{1+\frac{\sigma^2}{n-1} (\tilde  t-\tilde t_i) } } \\&
                        \tilde t_i=\frac{1}{ \frac{1}{\xi}\left (1+ \frac{\sigma^2}{n-1} ( \tilde t-\frac{\tilde t_i}{n})\right) +\frac{1}{1+\frac{\sigma^2}{n-1} (t-\frac{t_i}{n}) } .} 
                \end{split} 
        \end{align}
        Due to the symmetry between all users, $i=1,\ldots,n$, in (\ref{eq: t sec}) and the uniqueness of the solution, we have $t_i=t$  and $\tilde t_i=\tilde t$ which simplifies to:
        \begin{equation}\label{eq: t first}
        \begin{split} 
&       t=\frac{1}{ \frac{1}{\xi}\left (1+ \sigma^2 t\right) +\left(1+\sigma^2 \tilde  t \right)^{-1} } \\&
        \tilde t=\frac{1}{ \frac{1}{\xi}\left (1+ \sigma^2 \tilde t\right) +\left(1+\sigma^2 t \right)^{-1} } .
        \end{split} 
        \end{equation}
        Furthermore, using the uniqueness again, the symmetry between $t$ and $\tilde t$ gives $\tilde t=t$ and thus 
        \begin{equation}\label{eq: t second}
        t=\frac{1}{ \frac{1}{\xi}\left (1+ \sigma^2 t\right) +\left(1+\sigma^2 t \right)^{-1} } .
        \end{equation}
A simple manipulation on \eqref{eq: t second} yields the  cubic equation in (\ref{eq:cubic_main}). Further, due to the uniqueness property of  the theorem in Hachem et al. \cite{hachem2007deterministic}, the solution in (\ref{eq:cubic_main}) ensures a single unique root. This uniqueness property has been  further investigated  in  Section IV (see Corollary 2).
\end{IEEEproof}

In the next section, we use Theorem 1 to characterize the performance of the ZF and MMSE interference cancelers. Before that, we  briefly present a more general result that can help characterize  performance in non-homogeneous networks. Assume that all the elements of  matrix $\mathbf{Q}$ still have the same distribution, but can have a different variance. Define $\sigma^2_u=(M-1)\max[|q_{i,j}|^2]\big|, \forall i,j,i\neq j], $  and  $\sigma^2_l=(M-1)\min[|q_{i,j}|^2]\big|, \forall i,j,i\neq j]$ as the maximum and the minimum variance of an element in  the normalized FEXT matrix $\mathbf{Q}$, respectively. The following theorem provides a single parameter bounds on the $\tilde{\gamma}_n$.
\begin{my_theorem}
        If $\sigma_u^2<1$ and $\sigma_l^2<\sigma_u^2$ are the maximum and minimum of the absolute squared values of the normalized FEXT, respectively and  $t = \frac{1}{n} \text{tr}[{\mathbf T}(-\frac{1}{\xi})]$,  then 
\begin{align}\label{eq:theoreom_zf_lb}
\begin{split}
& \gamma_{\rm}^{\circ}\leq \frac{1}{1-\sigma^{2}_u} \\&
\lim_{\xi\to \infty}\gamma_{\rm}^{\circ}\geq\frac{1}{1-\sigma^{2}_l}
\end{split}
\end{align}
\end{my_theorem}

\begin{IEEEproof}
The proof is provided in Appendix A.
\end{IEEEproof}
Note that this theorem can also be used instead of Theorem 1 to characterize the performance of the ZF canceler in the homogeneous case ($\sigma_u^2=\sigma_l^2$).
\section{Performance of ZF and MMSE Cancelers}
In this section, we  use  Theorem \ref{theorem:main} to characterize the asymptotic  performance of  the  MMSE and ZF cancelers.
\subsection{Asymptotic Analysis of ZF Performance}
\begin{my_coro}[Deterministic equivalent of the associated average SNR loss]\label{lemma:zf}
        Let 
        \begin{equation} \label{eq:sigma_lemma_zf}
                \sigma^2=(M-1)\mathbb E[|q_{i,j}|^2]\big|_{i\ne j}.
        \end{equation}
        be the total FEXT power and the deterministic equivalent SNR loss given, respectively. If $\sigma^2<1$, as the system size grows to infinity, the average SNR loss of ZF will converge almost surely to the deterministic equivalent SNR loss,  $\gamma^o$:
        \begin{equation} \label{eq:th2_1}
                \lim_{n\to \infty} (\gamma^{\circ}-\tilde\gamma_n)=0 .
        \end{equation}
        where
                \begin{equation} \label{e:gamma_o}
        \gamma^{\circ}=(1-\sigma^2)^{-1}
        \end{equation}
        If $\sigma^2\ge 1$, then $\tilde \gamma_n$ is not bounded $(\lim_{n \rightarrow \infty}\tilde \gamma_n=\infty)$, i.e., the user rate will go to zero.
\end{my_coro}

\begin{IEEEproof}
Comparing  \eqref{eq:both_zf_mmse} with \eqref{eq:zf:trace} for the ZF, the deterministic equivalent of the SNR of the ZF canceler is given by 
by the root of \eqref{eq:cubic_main} when $\xi \to \infty$ such that
\begin{equation}\label{eq:zf_asymp}
\gamma^{ \rm ZF} \triangleq  \lim_{ \xi \to \infty } \gamma^{\circ}.
\end{equation}

If $t$ is bounded, we use $\lim_{ \xi \to \infty}$ in  \eqref{eq:cubic_main} to get

\begin{equation}\label{eq:zf_asymp2}
(1-\sigma^2)t-1 =0.
\end{equation}
which leads to \eqref{e:gamma_o}. However, if $t(z)$ is unbounded, we can rewrite equation  \eqref{eq:cubic_main} to get: \begin{IEEEeqnarray}{rCl}
                \sqrt{-\frac{1}{\xi}}t=\frac{1}{ -\left (\sqrt{-\frac{1}{\xi}}+ \sigma \sqrt{-\frac{1}{\xi}}t\right) +\left(\sqrt{-\frac{1}{\xi}}+\sigma \sqrt{-\frac{1}{\xi}} t \right)^{-1} } .
                \IEEEeqnarraynumspace
        \end{IEEEeqnarray}
        The unique solution of this equation when we take the limit as $\xi\to \infty$   is:
        \begin{equation}
        \lim_{\xi \to \infty} \sqrt{\frac{1}{\xi}}t=\frac{\sqrt{\sigma^2-1}}{{\sigma^2}}
        \end{equation}
        where a positive solution is obtained when $\sigma^2\ge 1$. Hence, if $\sigma^2>1$, $\sqrt{t/\xi}$ converges to a finite bound, and $t$ is not bounded.
\end{IEEEproof}
Thus, if $M$ is large enough and $\sigma^2<1$, $\gamma^{\circ}_{\rm}$ is a good approximation for $\tilde \gamma_n $. In this case, our asymptotic analysis provides a good approximation for the rate in the original system of interest. Compared to \eqref{eq:r_lb1}, and assuming that the Jensen inequality is tight, we conclude that the user rates in the system using the ZF canceler are well approximated by:
\begin{equation} \label{R:R approx_zf}
R_{i}^{\rm ZF}\simeq \mathbb E\bigg[{\log\Big(1+\frac{\eta_i}{\gamma_{\rm }^{\circ}}\Big)}\bigg], i=1\cdots M.
\end{equation}
By inspecting \eqref{R:R approx_zf}, this approximation is only a function of the number of users, $M$, and the statistics of the channel FEXT. Thus, the  result of Lemma \ref{lemma:zf} is very useful for system characterization, and can easily determine the regimes in which the ZF linear canceler is efficient and the regimes for which it is not a good choice.

As can be seen from \eqref{eq:sigma_lemma_zf}, for any system setup, the approximation will hold only up to a certain size $M$. This provides another illustration of the fact that \eqref{R:R approx_zf} will not necessarily be more  accurate for larger $M$. Obviously, this does not contradict Theorem \ref{theorem:main}, which is derived as $n$ (and not $M$) grows to infinity. Nevertheless, we need to provide an alternative intuition that will predict when \eqref{R:R approx_zf} is accurate. In Section  \ref{Sec:Numerical_s} we present a numerical study of the accuracy of  \eqref{R:R approx_zf}. We show that this approximation is very good for small values of $\sigma^2$, and holds well as long as $\sigma^2<0.5$.

The above Corollary \ref{lemma:zf}  determines that  asymptotic ZF performance is not useful when $\sigma^2\geq 1$. However, as the ZF and MMSE are equivalent at high SNR, the MMSE asymptotic result can be used to approximate the ZF performance when $\sigma^2\geq 1$ unlike the general performance bounds on these schemes where the ZF performance is used to predict  MMSE performance.

\subsection{Asymptotic Analysis of MMSE Performance}\label{subsec_mmse}
 By comparing \eqref{eq:both_zf_mmse} and \eqref{eq:gamma_general1},  the deterministic equivalent of $\gamma$ to compute the SNR loss for the MMSE can be  obtained directly from \eqref{eq:cubic_main} by substituting $\xi=\eta$.
However, for MMSE, it is more convenient to derive a direct approximation for the SNR, $\rho$ as defined in \eqref{eq:sinr_mmse}.
 
\begin{my_coro}[Deterministic equivalent of the MMSE SNR]\label{lemma:mmse}
        Let 
        \begin{equation} \label{eq:sigma_lemma_mmse}
        \sigma^2=(M-1)\mathbb E[|q_{i,j}|^2]\big|_{i\ne j}.
        \end{equation}
        and $\eta$ is the AWGN SNR, then the asymptotic SNR at the MMSE output is given by the positive root of
        \begin{align}
\label{eq:SPQ_mmse}
\begin{split}
        \rho^3-S\rho^2+Q\rho-P=0
\end{split}
\end{align}
where $ S=\eta-\eta\sigma^2-2$, $Q=1-2\eta$ and $P  =\eta^2\sigma^4+\eta\sigma^2+\eta$. The asymptotic SNR is given by the positive root:
        \begin{align}
\label{eq_:cubic_positive_mmse}
\begin{split}
\rho^{o}= -\frac{1}{3}\Big(-S+(-0.5+0.5\sqrt{3}i)^k\Delta+\\ \frac{S^2-3Q}{(-0.5+0.5\sqrt{3}i)^k\Delta}\Big), k =0, 1,2
\end{split}
\end{align}
where
$\Delta= \sqrt[3]{\frac{9SQ-2S^3-27P-\sqrt{(9SQ-2S^3-27P)^2-4(S^2-3Q)^3}}{2}}$.
\end{my_coro}

\begin{IEEEproof}
The cubic equation in \eqref{eq:SPQ_mmse} is  obtained directly from \eqref{eq:cubic_main} by substituting $\xi=\eta$ and $\rho=\eta/\gamma -1$ (see \eqref{eq:spectral_efficiency2}). The closed form solution in     \eqref{eq_:cubic_positive_mmse} can be obtained using Cardano's method.

To show that exactly one root of \eqref{eq:SPQ_mmse} is positive, we denote the three roots as $\rho_1$, $\rho_2$, and $\rho_3$. Note that these roots are either all real, or one root is real while the others form a complex conjugate pair. We observe that the three roots satisfy: $ P=\rho_1\rho_2 \rho_3$, $S=\rho_1+\rho_2+\rho_3$ and $Q=\rho_1 \rho_2+\rho_2\rho_3+\rho_3\rho_1$.
When $P$ is positive  the number of negative (real) roots is even. If there are 2 negative roots, the third is the only positive root and we are done. Thus, we need only  consider the case where there are no negative roots; i.e., either all roots are positive roots, or we have $1$ positive and $2$ complex roots. More specifically, we just need to rule out the case of all positive roots. This an be done if we either show that $S<0$ or  that $Q<0$.
Finally $Q<0$ if $\eta>0.5$, while $\eta\le 0.5$ guarantees that $S<0$. Thus, in all cases at least one  $S$ and $Q$ is negative, which rules out the possibility of all positive roots and completes the proof.
\end{IEEEproof}
Thus, if $M$ is large enough, $\gamma^{\circ}_{\rm}$  is a good approximation for $\tilde\gamma_n$, and thus $\rho^{\circ}$ in \eqref{eq_:cubic_positive_mmse} provides a good approximation  of the MMSE SNR.  As compared to \eqref{eq:spectral_efficiency2}, the user rates in the system are well approximated by:
\begin{equation} \label{R:R approx_mmse}
R^{\rm MMSE}\simeq {\log_2\Big(1+\rho^{\circ}_{\rm }\Big)}.
\end{equation}
Similar to the ZF, this approximation  is only a function of the number of users, $M$, and the statistics of the channel FEXT. Thus, the  result of Corollary \ref{lemma:mmse} is very useful for system characterization, and can easily determine the regimes in which linear cancelers are efficient and the regimes for which they are not a good choice. The asymptotic performance for the MMSE  can be  obtained for any  average FEXT value $\sigma^2$, whereas the asymptotic rate of the ZF canceler is zero for  $\sigma^2\geq1$.

 The asymptotic SNR using the MMSE canceler exhibits interesting and complicated behavior as a function of the SW-SNR and the FEXT power. To provide  additional insights, the following Corollary outlines the behavior of $\rho^{\circ}$ as a function of $\eta$ and $\sigma^2$. 

\begin{my_coro}
        \label{corr_charact}
 The behavior of the asymptotic SNR using the MMSE canceler as a function of $\eta$ and $\sigma^2$ can be characterized by:

\begin{enumerate}[label=(\alph*)]
        \item At low SW-SNR, $\eta$, the asymptotic SNR can be approximated by $\rho^{\circ}\approx (1+\sigma^2)\eta$.
        \item At high SW-SNR, $\eta$, the behavior of asymptotic SNR, depends on the total FEXT power:
\begin{itemize}
\item If $\sigma^2<1$, then $\rho^{\circ}\approx (1-\sigma^2)\eta$.
\item If $\sigma^2=1$, then $\rho^{\circ}\approx \sigma^4\eta^{2/3}$.
\item If $\sigma^2>1$, then $ \rho^{\circ}\approx \frac{\sigma^2}{\sqrt{\sigma^2-1}}\sqrt{\eta}$.
\end{itemize}

        \item If $\sigma^2$ is very small,  the SNR can be approximated by: $\rho^{\circ} = \eta+\eta\frac{ 1-\eta}{\eta   + 1}\sigma^2$.
        \item If  $\eta\le 1$, the SNR, $\rho^{\circ}$, increases monotonically with $\sigma^2$.
        \item If $\eta>1$ then:
        \begin{itemize}
                \item The SNR has   local minima with respect to $\sigma^2$.
                \item The location of the minima is at $0\le \sigma^2\le 1$ for $1<\eta<2.7$, at  $1\le \sigma^2\le 2$ for large $\eta$, and asymptotically approach $\sigma^2=2$ as $\eta\rightarrow\infty$.
        \end{itemize}
        \item For large $\sigma^2$ the SNR approaches $\sqrt{\eta \sigma^2}$.
\end{enumerate}

\end{my_coro}

This corollary leads to several interesting observations about the behavior of the MMSE receiver. Below, we first discuss this behavior and then give the proof of the corollary. 

Part $(a)$ is quite trivial. At low SNR, the interference is negligible, and the only issue is the total energy received relative to the Gaussian noise. 

Part $(b)$ presents the unique behavior of the MMSE receiver in the asymptotic regime. In any finite system at a high enough SNR, the performance of the MMSE converges to the performance of the ZF receiver. This indeed happens for $\sigma^2<1$. But, for $\sigma^2>1$, the ZF performance converges to $0$, while the MMSE is proportional to the square root of the SNR. The case of $\sigma^2=1$ lies on the border between the two other cases, but has a unique asymptotic behavior of its own.

It is interesting to note that unlike most results presented in this work, the convergence of the ZF performance to $0$ at $\sigma^2$ typically happens at very large system size. Thus, it was not observed in most of results in our numerical section. On the other hand, the similarity between MMSE and ZF at  high SNRs does hold even for quite large systems, and hence, for $\sigma^2>1$, we found that asymptotic MMSE performance gives a better prediction for the performance of ZF receivers at practical system sizes than  asymptotic ZF performance.

Part $(c)$  confirms that if $\sigma^2$ is low,  the direct channel is dominant. In the extreme case of  $\sigma^2=0$, the result $\rho^{\circ}=\eta$ is very intuitive, since the wireline channel matrix $\mathbf{H}_c$ becomes a diagonal and the  MMSE-SNR converges to the SW-SNR.

Part $(d)$  demonstrates (as in Part $(a)$) that for low SW-SNR the MMSE canceler can indeed use the FEXT, and the rate increases with $\sigma^2$ (although the low SNR regime is not practical in DSL).

Part $(e)$  illustrates the most unique characteristic of  MMSE performance as a function of the FEXT power ($\sigma^2$). The user SNR can either be monotonically decreasing, or have a single local minimum. In particular, in the high SNR which is more typical of most of wireline systems, the SNR will decrease for low FEXT powers but will eventually increase when the FEXT is large. 

 In Section  \ref{Sec:Numerical_s}, we present a numerical study of the accuracy of \eqref{eq:SPQ_mmse} and \eqref{R:R approx_mmse} for a general convergence analysis.  
\begin{IEEEproof} Although \eqref{eq_:cubic_positive_mmse} gives a closed form expression for $\rho^{\circ}$,in all the proofs we found it more convenient to start again with \eqref{eq:SPQ_mmse}.

Part $(a)$ is proved by substituting the low $\eta$ approximations into \eqref{eq:SPQ_mmse}: $S\approx -2$, $Q\approx1$ and $P  \approx \eta(1+\sigma^2)$, resulting in:
\begin{IEEEeqnarray}{rCl}
        \rho^3+2\rho^2+\rho\approx\eta(1+\sigma^2).
\end{IEEEeqnarray}
Realizing that $\rho$ will also be very small, we can  drop the higher powers of $\rho$, which leads directly to: $\rho\approx\eta(1+\sigma^2)$.

Part $(b)$ is proved through the same approach, using the high $\eta$ approximations: $ S\approx \eta(1-\sigma^2)$, $Q\approx -2\eta$ and $P  \approx \eta^2\sigma^4$, resulting in:
\begin{IEEEeqnarray}{rCl}
        \rho^3-\eta(1-\sigma^2)\rho^2 -2\eta\rho-\eta^2\sigma^4\approx0.
\end{IEEEeqnarray} 
Noting again that if $\eta$ is large $\rho$ will also be large, the third term will always be dominated by the other terms, and can be neglected:
\begin{IEEEeqnarray}{rCl}\label{e:high_eta_approx}
        \rho^3-\eta(1-\sigma^2)\rho^2 -\eta^2\sigma^4\approx0.
\end{IEEEeqnarray} 
Now, we need to distinguish between the three cases. When $\sigma^2<1$ the third term in \eqref{e:high_eta_approx} is negligible, leading directly to $\rho\approx\eta(1-\sigma^2)$. The proof for the second and third cases is equally simple, by noting that if $\sigma^2=1$, the second term in \eqref{e:high_eta_approx} vanishes, whereas if $\sigma^2>1$ the first term is negligible.

Part $(c)$ is proved by first noting that for $\sigma^2 = 0$, the
positive root of the cubic equation in \eqref{eq:SPQ_mmse} is  (as expected) $\rho^{\circ}=\eta$. Next, we 
evaluate the partial derivative $\rho^{\circ}$ with respect to $\sigma^2$, through the implicit derivative of Equation \eqref{eq:SPQ_mmse}:
        \begin{align}
        \label{eq:derivative}
        \begin{split}
        \rho^{'\circ}=\frac{d \rho^{\circ}}{d(\sigma^2)}& = \frac{ S' \rho^{2\circ}- Q'\rho^{\circ}+P'}{3 \rho^{2\circ}  - 2 S \rho^{\circ} + Q}\\&
=       \frac{ -\eta \rho^{2\circ}+2\eta^2\sigma^2+\eta}{3 \rho^{2\circ}  - 2 (\eta-\eta \sigma^2 -2) \rho^{\circ} + 1-2\eta}.
\end{split}
        \end{align}
For $\sigma^2=0$, and  also using $\rho^{\circ}=\eta$  in       \eqref{eq:derivative}, we get
        \begin{align}
\label{eq:derivative_sigma20}
\begin{split}
\rho^{'\circ}= \eta\frac{ 1-\eta}{\eta + 1}.
\end{split}
\end{align}
Thus, part $(c)$ of the Corollary is obtained as a first order Taylor approximation. Note that  part $(c)$ shows that if $\eta<1$, the first-order approximated  SNR, $\rho^{\circ}$, increases monotonically  with $\sigma^2$. However,  part  $(d)$  illustrates more stringent conditions on the monotonicity of the asymptotic SNR.  

Part $(d)$ is proved by finding the extremal points of $\rho^{\circ}$ as a function of $\sigma^2$. Using $\rho^{'\circ}=0$ in \eqref{eq:derivative} gives:
        \begin{align}
        \rho_*^2=2\eta\sigma_*^{2}+1
        \label{rh_star_sqr}
        \end{align}
where   $\sigma_*^{2}$ ans      $\rho_*$ are the extremal point of the average FEXT and the SNR value at that point, respectively. We use   \eqref{rh_star_sqr} in \eqref{eq:SPQ_mmse} to get the extremal point of the asymptotic SNR:
\begin{align}
\label{eq:rho_star_final}
        \rho_*=\frac{\eta^2\sigma_*^2(2-\sigma_*^2)+2\eta(1-2\sigma_*^2)-2}{2\eta(\sigma_*^2-1)+2}.
\end{align}
Recalling that  the SNR must be positive, we check when \eqref{eq:rho_star_final} is positive. We note that the numerator is positive only for:
\begin{align}
1-\frac{2}{\eta}-\sqrt{1-\frac{2}{\eta}+\frac{2}{\eta^2}}<\sigma_*^2<1-\frac{2}{\eta}+\sqrt{1-\frac{2}{\eta}+\frac{2}{\eta^2}}
\end{align}
while the denominator is positive only for:
\begin{align}
        \sigma_*^2>1-\frac{1}{\eta}.
\end{align}
By comparing the last two conditions, we conclude that  $\rho_*>0$ only for:
\begin{align}
\label{eq:range_sigma2}
        1-\frac{1}{\eta}<\sigma_*^2<1-\frac{2}{\eta}+\sqrt{1-\frac{2}{\eta}+\frac{2}{\eta^2}}.
\end{align}
The proof of Part $(d)$ is completed by noting that when $\eta<1$  \eqref{eq:range_sigma2} cannot be satisfied; hence, there is no extremal point of $\rho_*$. 

Part $(e)$ is proved by noting that \eqref{eq:range_sigma2} can be satisfied when  $\eta>1$, and evaluating the range of $\sigma^2$ for specific values of $\eta$. The asymptotic behavior for large $\eta$ is obtained by comparing the square root of \eqref{rh_star_sqr} with \eqref{eq:rho_star_final}, and taking the large $\eta$ asymptotic:
\begin{align}
\sqrt{2/\eta}
=\frac{\sqrt{\sigma_*^2}(2-\sigma_*^2)}{2(\sigma_*^2-1)}.
\end{align} 
As the left hand side goes to zero for large $\eta$, we must have $\sigma_*^2\rightarrow 2$. 

Part $(f)$  is proved in Theorem \ref{prop:wireless}, and discussed in the next subsection.
\end{IEEEproof}

\subsection{Comparison to wireless asymptotic results}
One natural question is how the  proposed analysis with assumptions AS1 and AS2  compares to the  results for wireless channels \cite{Liang2007_mmse}. In the wireless case, the SNR at the output of the MMSE receiver was shown to be (in terms of the parameters used in this paper):
\begin{align}\label{eq:wireless}
\rho_{\rm wireless}^{\circ}=-0.5+ 0.5\rm \sqrt{1+4\eta\sigma^2}.
\end{align}
However, this was derived for a channel matrix in which all elements are i.i.d., as opposed to the wireline channel matrix, in which the diagonal terms are fixed. To enable a comparison, we need to consider the case in which the effect of the diagonal elements on the SNR is negligible. This happens if we take $\sigma^2$ to infinity, and $\eta$ to zero while keeping their product constant (to maintain a finite output SNR). The following theorem proves that in this special case, the two solutions are identical.  As such our solution can be seen as a generalization of  \cite{Liang2007_mmse} to include both wireline and wireless channels.

\begin{my_theorem}
        \label{prop:wireless}
        When  $\sigma^2\to \infty$ and $\eta \to 0$  such that $\eta\sigma^2$ remains constant (i.e., the wireline channel approaches the wireless channel), then 
        $\rho^{\circ}\to-0.5+ 0.5\rm \sqrt{1+4\eta\sigma^2} =\rho_{\rm wireless}^{\circ} $ is the only positive root of (\ref{eq:SPQ_mmse}).
\end{my_theorem}
\begin{IEEEproof}
While we can work directly with the solution of \eqref{eq_:cubic_positive_mmse}, in this case, it is more convenient to go back to the cubic equation in \eqref{eq:SPQ_mmse}.         Using $\eta\to 0$ while  $\eta\sigma^2$ is  constant,  equation  (\ref{eq:SPQ_mmse}) can be written as:
        \begin{align}\label{eq_final2}
        (\rho+1)^3+(\eta\sigma^2-1)(\rho+1)^2-2\eta\sigma^2(\rho+1)-\eta^2\sigma^4=  0
        \end{align}
        Using long division, it can be seen that $\rho^{\circ}=-0.5+ 0.5\rm \sqrt{1+4\eta\sigma^2}$,   which exactly equal to the $\rho_{\rm wireless}^{\circ}$ in \eqref{eq:wireless}, is  one of the three roots of (\ref{eq_final2}). The other two roots are the solution of the quadratic equation:
        \begin{align}
        \label{eq:quad_dsl}
        \begin{split}
        (\rho+1)^2+(-0.5+\eta\sigma^2+0.5\sqrt{1+4\eta\sigma^2})(\rho+1)
        \\+\eta\sigma^2(-0.5+0.5\sqrt{1+4\eta\sigma^2}) = 0
        \end{split}
        \end{align}
        The solution of \eqref{eq:quad_dsl} yields the two roots: $\alpha=-1-\rho^{\circ}$ and $\beta=-1-\eta\sigma^2$. Since $\rho^{0}$, $\eta$ and $\sigma^2$ are positive,  the cubic equation (\ref{eq_final2}) has a single  positive root and two negative roots.  
\end{IEEEproof}

Theorem \ref{prop:wireless} confirms that the wireless solution \cite{Liang2007_mmse}  is a special case of the generalized solution proposed in this paper. Thus the proposed asymptotic analysis  performs better for wireline channels than the result developed for  wireless channels \cite{Liang2007_mmse} and is expected to converge excellently at lower values of SNR and $\sigma^2$.  Hence, we expect that both  solutions will yield similar results when the diagonal elements of the channel matrix are similar to the non-diagonal elements. However, the proposed asymptotic result should perform better for  typical wireline channels where the diagonal elements are distributed independently to the non-diagonal elements.

\section{Numerical and Simulation Analysis}\label{Sec:Numerical_s}

\begin{table}[t]
        \caption{Simulation Parameters for DSL Wireline Systems}
        \label{simulation_parameters}
        \centering
        \begin{tabular}{cc}
                %\hline
                %\bfseries First & \bfseries Next\\
                        \hline
                Number of users  & $M$ = $10$ to  $100$\\
                \hline
                Band plan (G.fast) & $106$ \mbox{MHz} and $212$ \mbox{MHz}\\
                \hline
                Band plan (VDSL) & $30$ MHz\\
        
                        \hline
                Tone spacing (VDSL)  &  $\Delta_f$ = $4.3125$ \mbox{KHz}\\
                \hline
                        Tone spacing (G.fast)  &  $\Delta_f$ = $51.75$ \mbox{KHz}\\
                \hline
                Signal PSD $f\leq 30 $ \mbox{MHz} & $-65$ \mbox{dBm/Hz}   \\
                \hline
                Signal PSD $30<f\leq 106$ {MHz} & $-76$ \mbox{dBm/Hz}   \\
                \hline
                Signal PSD $f> 106$ {MHz} & $-79$ \mbox{dBm/Hz}   \\
                \hline
                Additive noise & $-140$ \mbox{dBm/Hz}\\
                \hline
                        Noise margin & $6$ \mbox{dB}\\
                \hline
                Coding gain & $5$ \mbox{dB}\\
                \hline
                Target BER & $10^{-7}$\\
                \hline
                Shannon Gap $\Gamma$ & $9.75$ \mbox{dB}\\
                \hline
        bit cap (VDSL) & $15$ bits\\
        \hline  
        bit cap (G.fast) & $12$ bits\\
        \hline
        \end{tabular}
\end{table}

In this section, we study the convergence of the actual SNR to the asymptotic analysis of linear cancelers for  wireline systems through computer  simulations.   First, we consider a general wireline channel model, and then consider the G.fast \cite {g9701} and VDSL \cite{g993} wireline standards as an example to demonstrate the proposed analysis.

We start with a simplified scenario by generating  general random channel matrices $\mathbf{Q}$ that satisfy the assumptions AS\ref{assump:one} and AS\ref{assump:two}. Here, the diagonal elements of  matrix $\mathbf{Q}$ are zero, while the non-diagonal elements are i.i.d and log-normally distributed with a mean and variance selected from  \cite{sorbara2007construction}. In these simulations we used  the zero, low and high coupling models of  \cite{sorbara2007construction} to simulate a variety of wireline channels.

\begin{figure}[t]
        \centering
        {\includegraphics[scale=0.45]{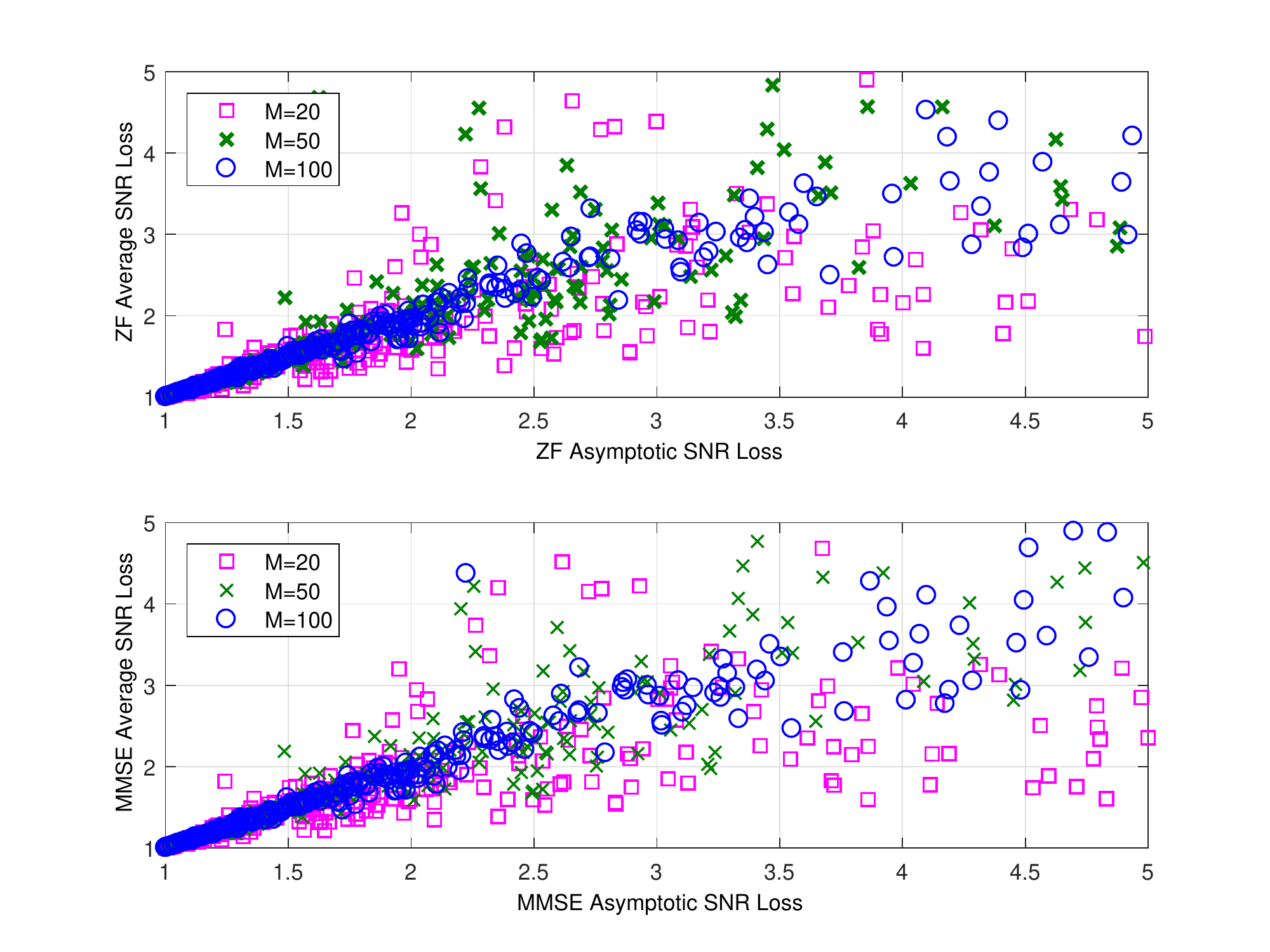}}
        \caption{Asymptotic versus average SNR  loss of ZF and MMSE cancelers for $1000$ channel realizations for matrix size $M\in\{20,50,100\}$ with $0\leq\sigma^2< 1$. Single wire SNR is $30$ \mbox{dB}.}
        \label{fig:perf_scatter_var1}
 \end{figure}

To demonstrate the accuracy of the suggested approximation in \eqref{R:R approx_zf} and  \eqref{R:R approx_mmse}, we evaluated the average SNR loss of  ZF and MMSE receivers  for $1000$ different channel setups.  The average SNR loss in each setup is presented in Fig. \ref{fig:perf_scatter_var1}, as a function of the asymptotic expression. As can be seen, regardless of the matrix size ($M$), the accuracy of \eqref{R:R approx_zf} and \eqref{R:R approx_mmse} is very good as long as the average SNR loss is below $1$. When solving $\sigma^2$ from \eqref{e:gamma_o} for the ZF canceler,  the approximation turns out to be quite accurate as long as the total FEXT power, $\sigma^2$ is less than half of the power of the direct channel which is typically observed for DSL channels. 

\begin{figure}[t]
        \centering
        {\includegraphics[scale=0.45]{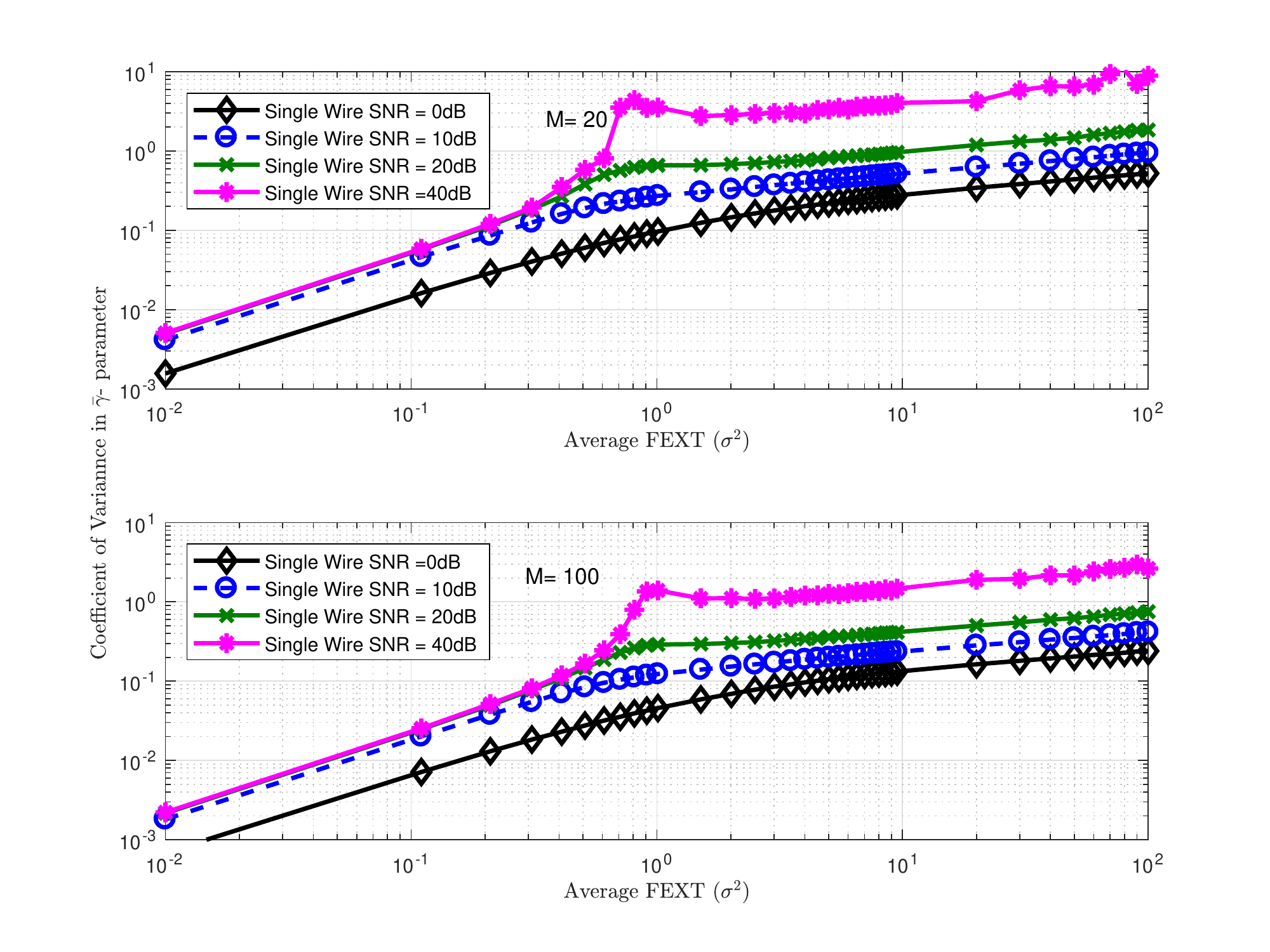}}
        \caption{Coefficient of variation  of the $\gamma$-parameter of the MMSE canceler for channel matrix size $M=20$ and $M=100$ at various single wire SNRs.}
        \label{fig:perf_scatter_var2}
 \end{figure}

 In contrast  to the ZF canceler approximation which approaches  zero, the  approximation for the MMSE canceler  converges for non-zero values even for $\sigma^2>1$, and  also depends  on the single-wire SNR.  To study the performance of the MMSE,  Fig.  \ref{fig:perf_scatter_var2} depicts the variation  of the approximation error  in the $\bar{\gamma}$-parameter by considering channels from low  coupling models in  \cite{sorbara2007construction}. The coefficient of variation of the $\gamma$-parameter is defined as its standard deviation divided by the mean; i.e., $\frac{\sqrt{\mathbb{E}[|\bar{\gamma}-\gamma^{\circ}|^2]}}{\mathbb{E}[\bar{\gamma}]}$, where $\bar{\gamma}$  was given in \eqref{eq:gamma_general}. It can be seen that the MMSE approximation improves at lower  SW-SNR and lower values of average FEXT ($\sigma^2$) and large matrix size, $M$.

\begin{figure}[t]
        \centering
        {\includegraphics[scale=0.45]{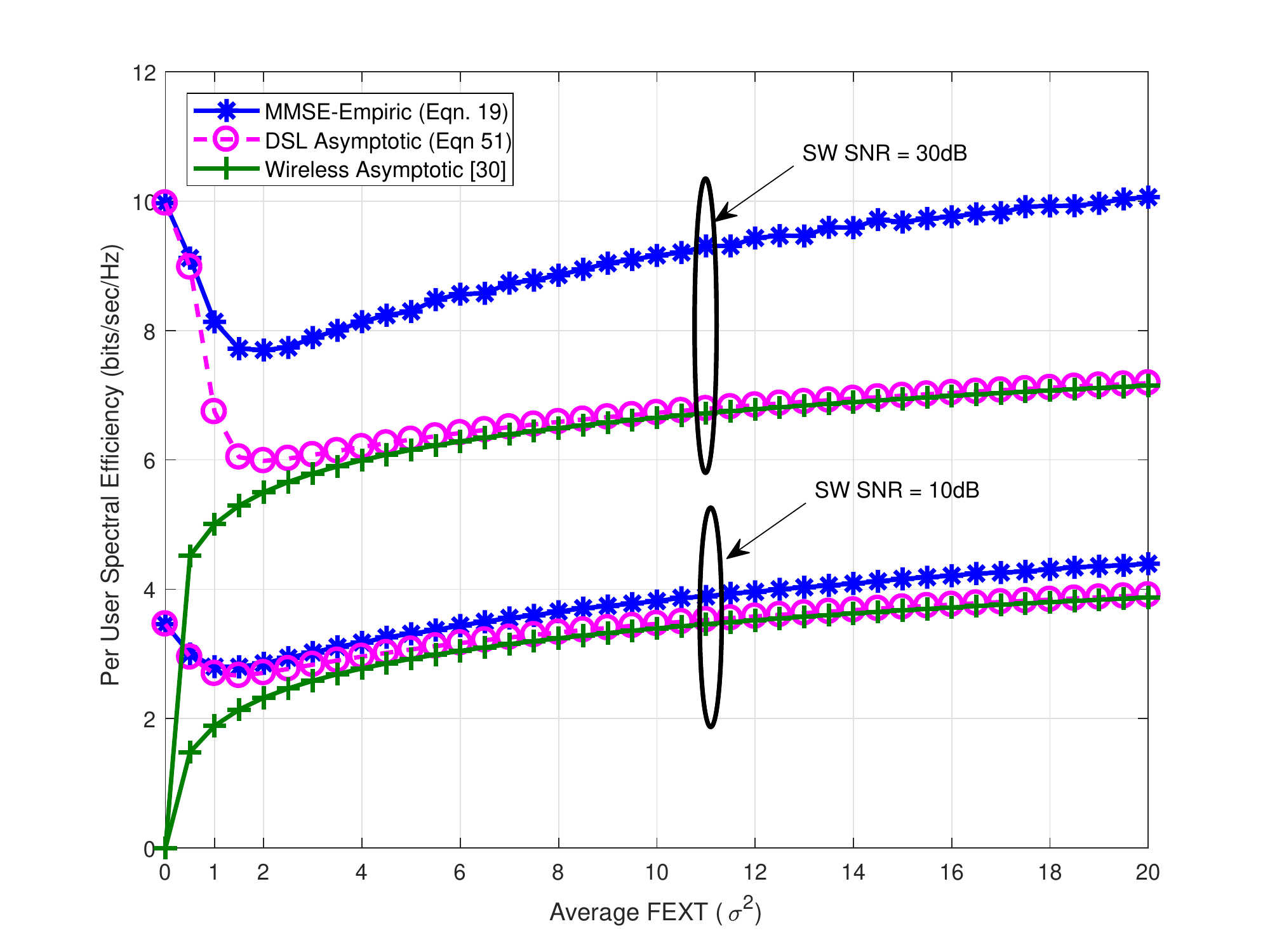}}
        \caption{Performance of the wireline asymptotic analysis  compared to the wireless asymptotic for the MMSE canceler for a general random  channel matrix $\mathbf{Q}$ of size  $M=20$. The SW-SNR is set to $\eta=1$ and  non-diagonal elements  are i.i.d and log-normally distributed.}
        \label{fig:dsl_vs_wireless}
\end{figure}

In order to further  demonstrate the significance of the proposed asymptotic analysis for  wireline systems, we compared it to the wireless asymptotic result in \cite{Liang2007_mmse}. In contrast to the wireline  channel whose diagonal elements (direct path for signal transmission) differ from the non-diagonal elements (crosstalk paths), the wireless analysis assumes that all the elements of  channel matrix are identically distributed  (due to random multi-paths\footnote{Even in line of sight scenarios, the wireline channel differs from the wireless channel since the wireless channel will have the same statistics for all antennas (each antenna will see a line of sight component and a fading component). In contrast, in  wireline, the diagonal element has a specific characteristic, 
because it is the only element with a direct wire connection.}) \cite{Liang2007_mmse}.    Fig.~\ref{fig:dsl_vs_wireless} compares the asymptotic approximation of the average spectral efficiency using the  technique of wireless system \cite{Liang2007_mmse}, the proposed analysis in  this paper, and the simulation results, for various average FEXT ($\sigma^2$) and SW-SNR $\eta$.  The figure shows that that the approximation using the  proposed wireline asymptotic performs quite well for  the whole range of $\sigma^2$ for lower values of SW-SNR ($10$ \mbox{dB}), but for higher SW-SNR ($30$ \mbox{dB}) it is accurate only up to $\sigma^2<0.5$. Note that these are typical values observed in DSL systems. On the other hand, the wireless asymptotic completely fails to predict the output SNR  at low $\sigma^2$, where the effect of the diagonal elements is dominant (the wireless asymptotic actually converges to zero for $\sigma^2\rightarrow 0$). Both approximations show similar behavior at higher FEXT ($\sigma^2$), where the effect of diagonal elements becomes negligible compared to the non-diagonal elements.

To test a more practical scenario, we analyzed the approximation of the performance of linear cancelers over measured channels and stochastic channel models of DSL systems (VDSL and G.fast).  The measured channels for VDSL  contained data from a $26$ AWG cables with $28$ pairs of various line lengths  \cite{U-BROAD}. We considered  measured channels for G.fast from  a $0.5$ \mbox{mm} BT cable with $10$ pairs each measuring $100$ \mbox{m} \cite{BT}. 
 The simulated MIMO channels were obtained using  parametric DSL models (CAD55 cable) at various line lengths and binder size \cite{g9701} \cite{sorbara2007construction}.  The other parameters used in the simulations are listed in Table \ref{simulation_parameters}.

\begin{figure}[t]
        \centering
        {\includegraphics[scale=0.45]{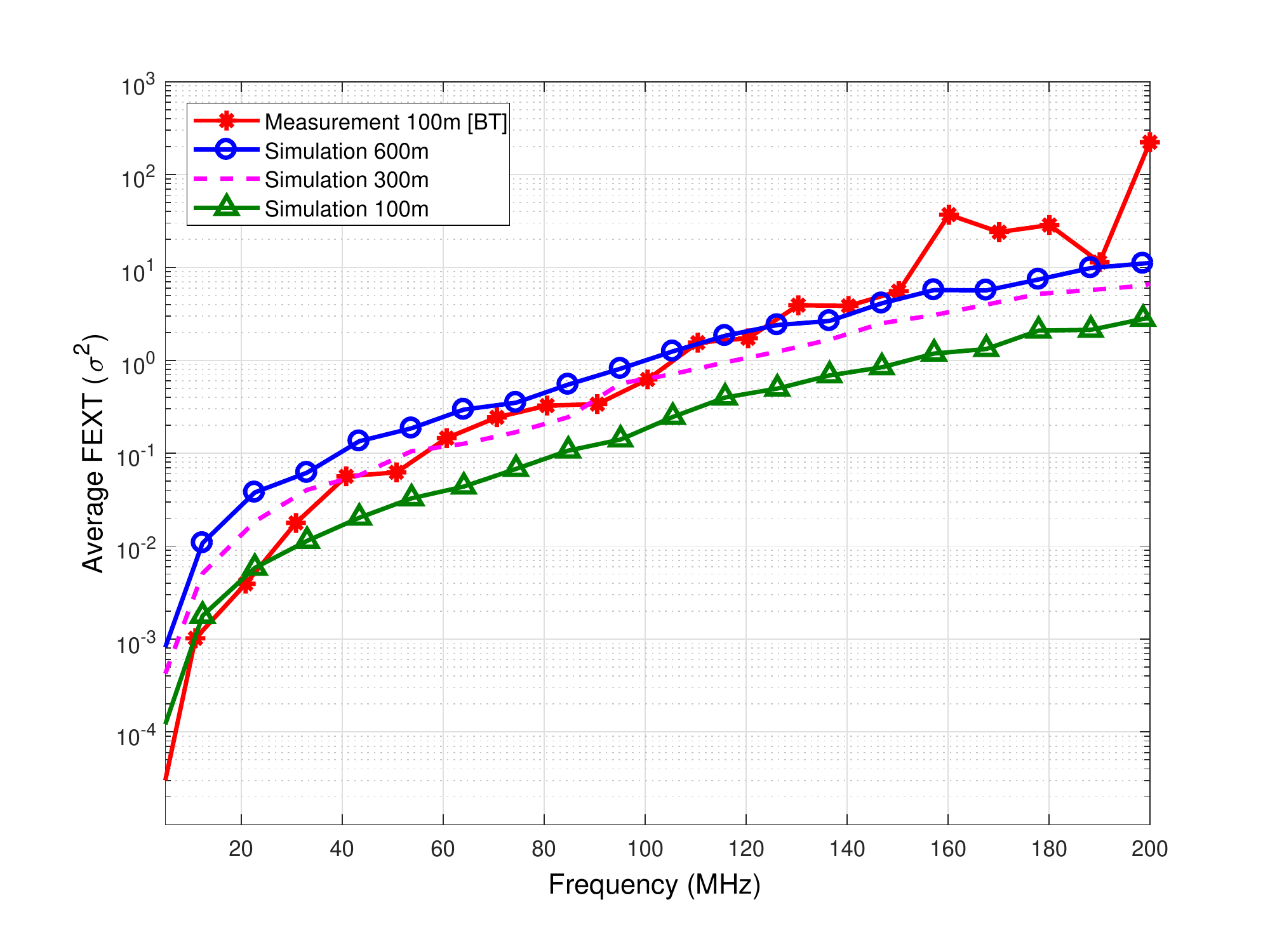}}
        \caption{Average FEXT i. e., $\sigma^2$  of DSL measurement channel matrix \cite{BT}   and simulated channel matrix  (cable CAD55) of size $M=10$ for G.fast frequencies.}
        \label{fig:avg_fext}
\end{figure}

An important step was to validate the  requisite  assumptions AS\ref{assump:one} and AS\ref{assump:two} from the measured  DSL channels. The random behavior of the FEXT and the statistical characterization of the channel were  studied   in the conference version of this paper \cite{zafaruddin_eilat2016}\footnote{The DSL channel is considered typically stationary; however, the FEXT coefficients are random  which increases with frequencies.}. Since the total average FEXT per user $\sigma^2$ is an important quantity for the accuracy of approximation, we present the average FEXT values of a few channels in  Fig. \ref   {fig:avg_fext}. The figure shows that the average FEXT values increase with frequency and can be very high ($\sigma^2>100$) at very high frequencies. However, at such  high frequencies, the SNR is quite low (which helps  convergence) and  the spectral efficiency is also quite low hence  the impact on the overall data rate is small. 

 \begin{figure}[t]
        \centering
        {\includegraphics[scale=0.45]{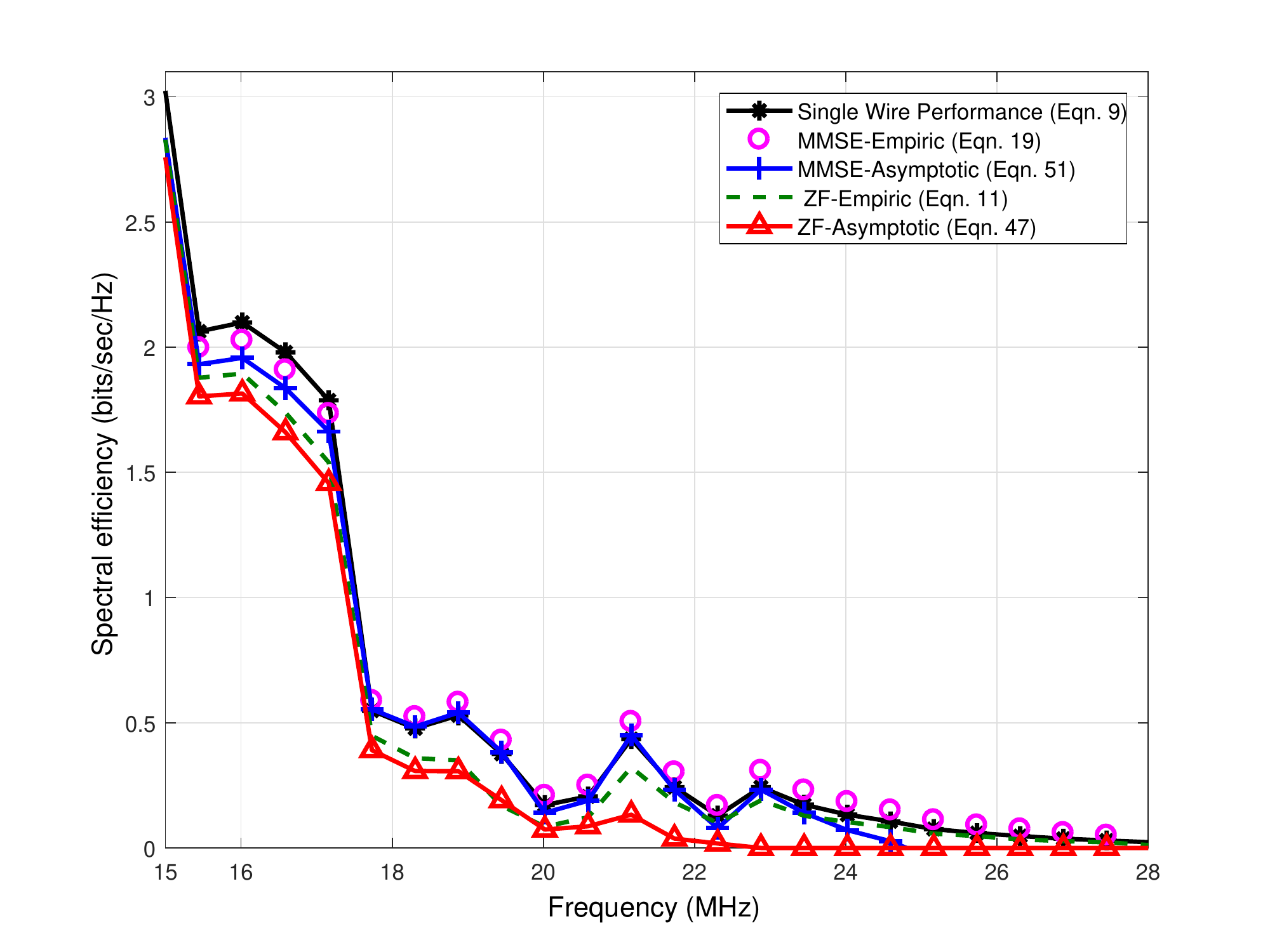}}
        \caption{Performance of asymptotic analysis (spectral efficiency vs. frequency of transmission) of measurement channels for  VDSL frequencies with $M=28$, loop length of $590$m, and cable type $26$ AWG.}
        \label{fig:avg_perf_vdsl}
\end{figure}

\begin{figure}[t]
        \centering
        {\includegraphics[scale=0.45]{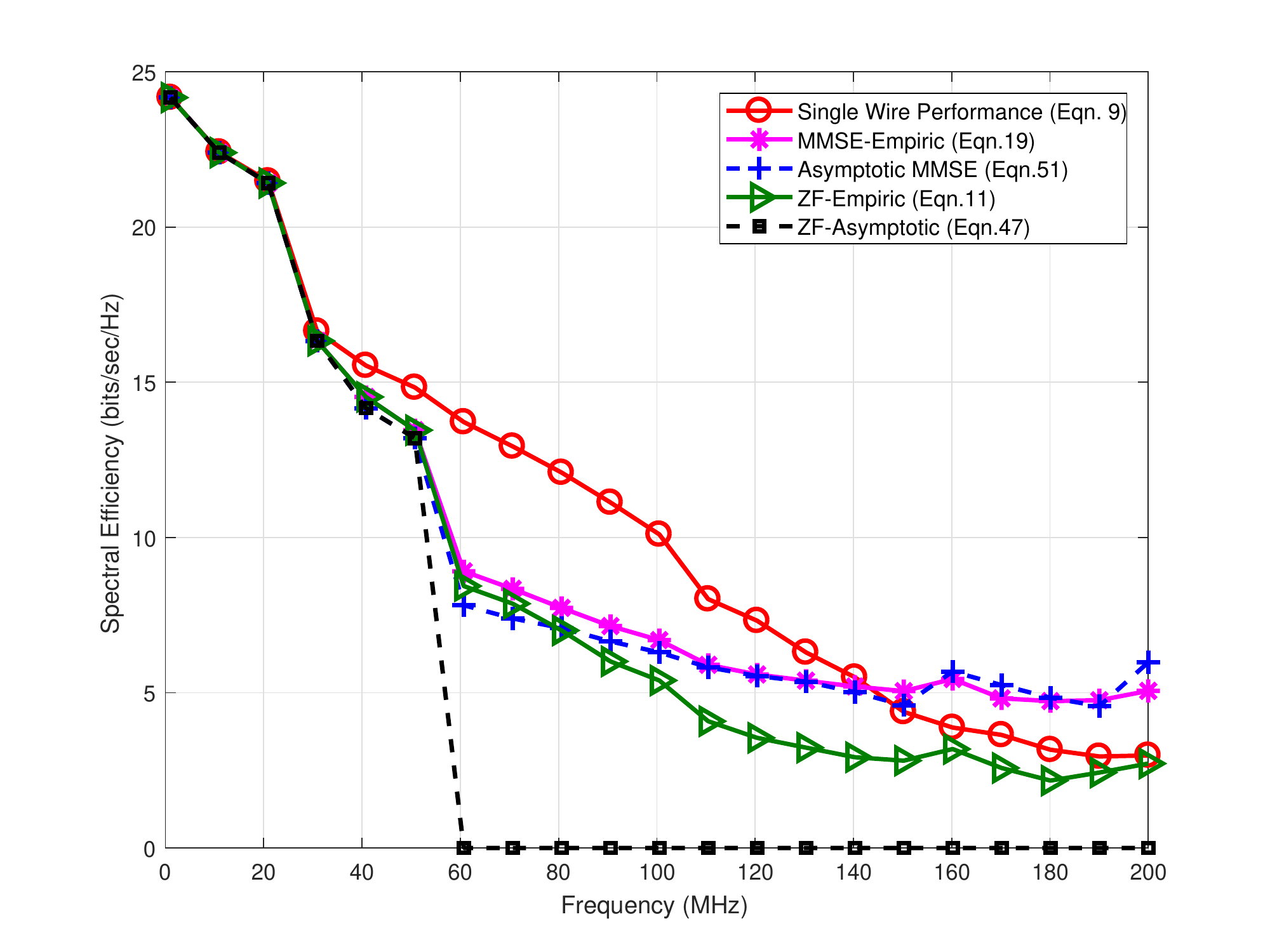}}
        \caption{Performance of asymptotic analysis (spectral efficiency vs. frequency of transmission) of measurement channels for G.fast frequencies with measurement data for $100$ pairs obtained using $10$ pair data \cite{BT}.}
        \label{fig:avg_perf_gfast}
\end{figure}

%For the G.fast $106$ \mbox{MHz} with a $10$ users binder, the simulations in Fig.~ \ref{fig:spectral_efficiency}b show that the approximation for both cancelers is very accurate for most of the except at very high frequency tone where it becomes less tight since the average loss of the cancelers is more than $3$ \mbox{dB} (i.e. when the SNR loss is greater than $2$). Also the MMSE performs better than the ZF canceler  when $\sigma^2$  is close to $1$ and when  $\sigma^2>1$  where  the ZF canceler converges to the zero-rate. 

 Figures \ref   {fig:avg_perf_vdsl} and \ref    {fig:avg_perf_gfast} depict the performance of the ZF and MMSE cancelers over measured channels (for both VDSL and G.fast frequencies). The plots show the  spectral efficiency per user (calculated by averaging (\ref{eq:rr}) for the ZF and \eqref{eq:spectral_efficiency} for the MMSE)  and the asymptotic approximation for the ZF canceler $R^{\rm ZF}$  in \eqref{R:R approx_zf} and $R^{\rm MMSE}$ in \eqref{R:R approx_mmse}  for the MMSE. For reference, we also depict the spectral efficiency of a single wire (setting $ \bf H = I$) such that only a single wire is active in a binder without the effect of crosstalk. For Fig. \ref      {fig:avg_perf_gfast}, we constructed a $100$ user channel matrix by concatenating a randomly permuted version of the measured matrix in \cite{BT}. Both figures  show that  the approximation is quite accurate for both the ZF and the MMSE cancelers for VDSL channels, even for longer loops ($\sigma^2$ increases with loop length). This takes place because of the smaller average FEXT, $\sigma^2$, as shown in Fig.~\ref     {fig:avg_fext}. On the one hand, Fig.  \ref     {fig:avg_perf_gfast} shows  that the MMSE asymptotic result is quite accurate but the ZF asymptotic goes to zero above $60$MHz. Note that the zero asymptotic results shows that the performance will deteriorate as the system size increases. Nevertheless, a zero is never a good approximation;  hence,  the ZF result is not useful for $\sigma^2\ge 1$. Nevertheless, at these SNRs, the ZF performs quite close to the MMSE, and the MMSE asymptotic result gives quite a good approximation for both.

\begin{figure}[t]
        \centering
        {\includegraphics[scale=0.45]{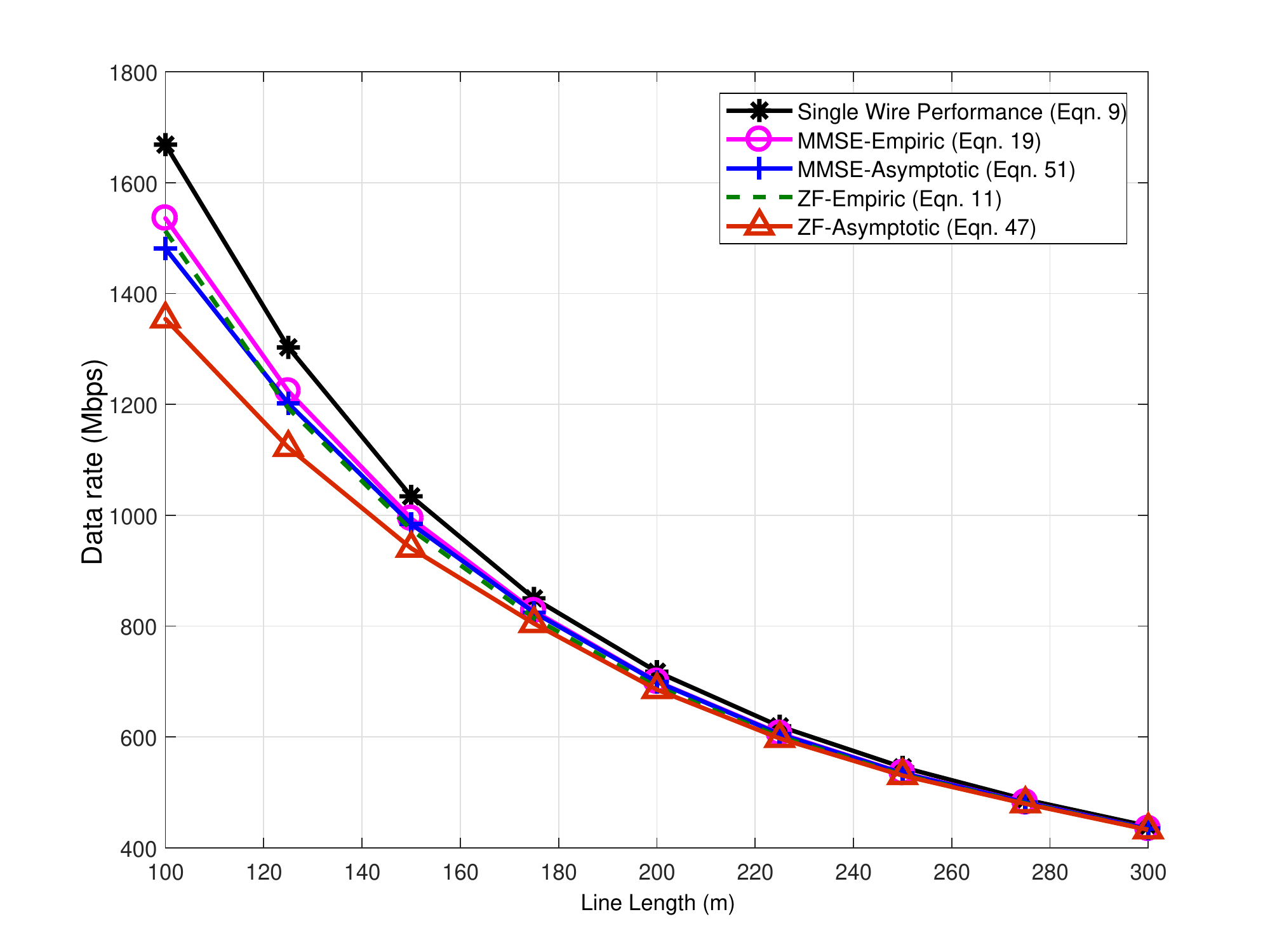}}
        \caption{Average  user rate performance of the asymptotic analysis (rate versus reach) using parametric DSL channels (CAD55) for G.fast $212$ \mbox{MHz}. The binder is composed of $20$ users with equal line lengths.}
        \label{fig:avg_data_rate}
\end{figure}

Finally, the performance is reported in rate versus reach curves for parametric CAD55 DSL channels by incorporating the Shannon-gap and the bit-cap from  Table \ref{simulation_parameters}. The aggregated rates were obtained by summation of spectral efficiency of the ZF canceler in (\ref{eq:rr}) and (\ref{R:R approx_zf}), and \eqref{eq:spectral_efficiency}  and \eqref{R:R approx_mmse} for the MMSE  performance, over all DMT tones with a tone width of $51.75$ \mbox{KHz}. The rate-reach curves compared average data rates at a specific line length range for various channel realizations with that of the asymptotic approximation. Fig.~\ref{fig:avg_data_rate} depicts the accuracy of approximation over  $212$ \mbox{MHz} G.fast systems. It shows that the approximation is quite accurate even  for the G.fast system.  However, the MMSE result is more accurate in predicting the performance of both the MMSE and the ZF cancelers.  Since  DSL systems operate at a very high SNR, the asymptotic analysis of the MMSE can also be used for ZF performance.

 \section{Conclusion}
In this work, we presented an asymptotic analysis of the performance of ZF and MMSE receivers in a large matrix wireline system. We derived an approximation for the user rate for both the ZF and MMSE cancelers which are very simple to evaluate and does not need to take specific channel realizations into account. The analysis was based on the theory of large dimensional random matrices, but  computer simulations over measured and simulated DSL channels showed that this approximation was accurate even for relatively low dimensional systems. We showed  that the proposed asymptotic analysis  converges excellently for relatively low FEXT or low SNR,   and is thus useful for most scenarios in practical DSL systems.   
 \section*{Appendix A: Proof of Theorem 2}
        Substituting $\mathbf \Psi(-\frac{1}{\xi})$ and         $\tilde{\mathbf \Psi}(-\frac{1}{\xi})$ from (\ref{eq:Psi(z)}), the diagonal matrix ${\mathbf T}(-\frac{1}{\xi})$ can be represented as
 \begin{align}\label{eq:simplify}
 {\mathbf T}(-\frac{1}{\xi}) =diag\Big (\frac{1}{{\psi}_1^{-1}(-\frac{1}{\xi})+\frac{1}{\xi}\tilde{\psi}_1(z)},\ldots,\nonumber \\ \frac{1}{{\psi}_n^{-1}(-\frac{1}{\xi})+\frac{1}{\xi}\tilde{\psi}_n(-\frac{1}{\xi})}\Big), 
 \end{align}
 Hence,  the trace of ${\mathbf T}(-\frac{1}{\xi})$:
 \begin{equation}\label{eq:trace}
 \text{tr}[{\mathbf T}(-\frac{1}{\xi})]= \sum_{k=1}^n\frac{1}{{\psi}_k^{-1}(-\frac{1}{\xi})+\frac{1}{\xi}\tilde{\psi}_k(-\frac{1}{\xi})}.
 \end{equation}
 Now, we        substitute  $\psi_k$ and $\tilde{\psi}_k$ from  (\ref{eq:psi(z)}) in (\ref{eq:trace}) to get:
 \begin{equation}\label{eq:trace_one1}
 \text{tr}[{\mathbf T}(-\frac{1}{\xi})]= \sum_{k=1}^n\frac{1}{\frac{1}{\xi}\left(1+\text{tr}\big(\tilde{\mathbf \Omega}_k\tilde{\mathbf T}(-\frac{1}{\xi})\big)/n\right)+\frac{1}{1+\text{tr}\left({\mathbf \Omega}_k{\mathbf T}(-\frac{1}{\xi})\right)/n}}
 \end{equation}
 First, we prove the upper bound.       Since $\xi\to \infty$ and $\text{tr}\big(\tilde{\mathbf \Omega}_k\tilde{\mathbf T}(-\frac{1}{\xi})\big)/n\geq 0$, \eqref{eq:trace_one1} yields an upper bound on $\text{tr}[{\mathbf T}(-\frac{1}{\xi})]$ as
 \begin{align}\label{eq:trace_one2}
 \begin{split}
 \text{tr}[{\mathbf T}(-\frac{1}{\xi})]\leq  \sum_{k=1}^n 1+\sum_{k=1}^n\frac{1}{n}\text{tr}\left({\mathbf \Omega}_k{\mathbf T}(-\frac{1}{\xi})\right)\\
 \leq n+ \text{tr}\left(\frac{1}{n}\sum_{k=1}^n {\mathbf \Omega}_k{\mathbf T}(-\frac{1}{\xi})\right),
 \end{split}
 \end{align}    
 where we used the identity $\text{tr}(\mathbf{A}+\mathbf{B}) =         \text{tr}(\mathbf{A})+\text{tr}(\mathbf{B})$ in the first term for $k=1 \cdots n$. 
 Using $\frac{1}{n} \sum_{k=1}^n{\mathbf \Omega}_k\leq \sigma_u \mathbf{I} $ from (\ref{eq:D_i})  and  $t = \frac{1}{n} \text{tr}[{\mathbf T}(-\frac{1}{\xi})]$, (\ref{eq:trace_one2}) can be simplified to get $t \leq \frac{1}{1-\sigma_u^2 }$.     Similarly, we can analyze $\tilde{\mathbf T}(-\frac{1}{\xi})$  to prove that      $\tilde {t}\leq \frac{1}{1-\sigma_u^2}$. Due to the uniqueness of the solution, and applying $\lim_{\xi \to \infty}$ to $t$ or $\tilde{t}$, we get the upper bound in \eqref{eq:theoreom_zf_lb}.

  For the lower bound, we apply $\lim_{z\to 0^{-}}$ to (\ref{eq:trace_one1}):
 \begin{align}\label{eq:trace_one6}
 \lim_{\xi \to \infty} \text{tr}[{\mathbf T}(-\frac{1}{\xi})]=n+\text{tr}\left(\frac{1}{n}\sum_{k=1}^n {\mathbf \Omega}_k{\mathbf T}(-\frac{1}{\xi})\right),
 \end{align}
 where we  used the fact that $\tilde{\mathbf \Omega}_k\tilde{\mathbf T}(-\frac{1}{\xi})$ is bounded for every $k$.
 Using $\frac{1}{n} \sum_{k=1}^n{\mathbf D}_k\geq \sigma_l \mathbf{I}$ from (\ref{eq:D_i})  and $t = \frac{1}{n} \text{tr}[{\mathbf T}(-\frac{1}{\xi})]$ in (\ref{eq:trace_one6}), we get        $ \gamma_{\rm}^{\circ} \geq \frac{1}{1-\sigma_l^2 }$.     Similarly, using $\tilde{\mathbf T}(-\frac{1}{\xi})$, we can have $\gamma_{\rm}^{\circ} \geq \frac{1}{1-\sigma_l^2 }$, hence the lower bound in \eqref{eq:theoreom_zf_lb}.

\bibliographystyle{ieeetran}
\bibliography{asymptotic_analysis_of_DSL_systems_letter}

\end{document}